%% file: final.tex
\documentclass[a4paper,12pt]{article}
\usepackage[dvips]{epsfig}
\usepackage{rotating}

\newcommand{\xe}[0]{x_{b}}
\newcommand{\xeb}[0]{x_B}
\newcommand{\xel}[0]{x_B^{\mathrm L}}
\newcommand{\xewd}[0]{x_B^{\mathrm {wd}}}
\newcommand{\xerec}[0]{x_B^{\mathrm{obs}}}

\newcommand{\mxel}[0]{\langle x_B^{\mathrm L} \rangle }
\newcommand{\mxeb}[0]{\langle x_B \rangle }
\newcommand{\mxewd}[0]{\langle x_B^{\mathrm {wd}} \rangle }
\newcommand{\etal}[0]{{\it{et al.}}}
\newcommand{\pisoft}[0]{\pi_{\mathrm s}}
\newcommand{\udsc}[0]{non-$b\bar b$}
\newcommand{\dedx}[0]{\mathrm{d}E/\mathrm{d}x}

\begin{document}
\begin{titlepage}
\pagestyle{empty}
%

\begin{center}
EUROPEAN ORGANIZATION FOR NUCLEAR RESEARCH (CERN)
\end{center}

\vspace{1 cm}
\begin{flushright}
  CERN-EP/2001-039\\
May 21st, 2001
\end{flushright}

\vspace{1 cm}
\begin{center}
\boldmath
{\LARGE\bf 
Study of the 
fragmentation of $b$ quarks into $B$ mesons at the $Z$ peak
}
\vskip 1cm
{\Large The ALEPH Collaboration$^{*)}$}

\end{center}
\vskip 1.5 cm \centerline{\large\bf Abstract} \vskip .7cm {\noindent
  The fragmentation of $b$ quarks into $B$ mesons is studied with four
  million hadronic $Z$ decays collected by the ALEPH experiment during
  the years 1991--1995.  A semi-exclusive reconstruction of $B\to \ell
  \nu D^{(\star)}$ decays is performed, by combining lepton candidates
  with fully reconstructed $D^{(\star)}$ mesons while the neutrino
  energy is estimated from the missing energy of the event.
  
  The mean value of $\xewd$, the energy of the weakly-decaying $B$
  meson normalised to the beam energy, is found to be
\begin{displaymath}
  \mxewd = 0.716 \; \pm 0.006 \, (\mathrm{stat}) \; \pm 0.006 \, (\mathrm{syst}) 
\end{displaymath}
using a model-independent method; the corresponding value for the
energy of the leading $B$ meson is $\mxel = 0.736 \; \pm 0.006 \,
(\mathrm{stat}) \; \pm 0.006 \, (\mathrm{syst})$. The reconstructed
spectra are compared with different fragmentation models.  }

\vspace{4 cm}
\begin{center}
\em  To be submitted to Physics Letters
\end{center}

\vspace{2 cm}
*) See next pages for the list of authors

\end{titlepage}

\newpage
{\include{authors}}
\newpage

\section{Introduction}
The process of hadron production at $e^+ e^-$ colliders is usually
modelled as the convolution of a perturbative part (hard gluon
radiation for energies above approximately 1~GeV) and a
non-perturbative part, called hadronisation or fragmentation, in which
the quarks are confined in colourless hadrons.  While the first step is
in principle calculable, the fragmentation needs a
phenomenological approach and is usually parametrised in terms of the
variable
\begin{equation}
z\equiv  \frac{\left( E+ p_{\|}\right)_{\mathrm{hadron}}}{ \left( E+ p\right)_{\mathrm{quark}}}  \quad ,
\end{equation}
where $p_{\|}$ is the hadron's momentum along the direction of the
quark, and $\left( E+ p\right)_{\mathrm{quark}}$ is the sum of the
quark energy and momentum just before fragmentation, i.e. taking into
account initial and final state photon radiation, and hard gluon emission.

With this definition, the fragmentation process can be described in
terms of the probability of a hadron $H$ to be generated with a given $z$,
called $D^H_q(z)$, where $q$ is the flavour of the generating quark.
In this paper the fragmentation
of  $b$ quarks is studied.

The fraction $z$  is not accessible experimentally,
and hence a direct reconstruction of $D^H_b(z)$ is not possible.
The energy spectrum of $b$ hadrons is therefore described in terms 
of the scaled energy $\xe$, 
defined as the ratio of the heavy hadron energy to the beam energy
\begin{equation}
  \xe \equiv \frac{E_\mathrm{had}}{E_{\mathrm{beam}}} \quad .
\end{equation}
In contrast to the $z$ variable, the effects of initial and final
state radiation and hard gluon emission are not unfolded.

In the analysis presented, the energy of $B$ mesons is
reconstructed using a partially exclusive method: semileptonic decays
$B\to \ell \nu D^{(\star)}$ are identified by pairing lepton
candidates with fully reconstructed $D^{(\star)}$ mesons; the scaled
energy of the weakly-decaying $B$ meson is then computed
adding an estimate of the neutrino energy.
Five channels are chosen because of their good signal purity 
and statistical significance; 
they are shown in Table~\ref{tab:channels}. 

In the following $\xerec$ indicates the reconstructed energy of $B$
meson candidates, $\xewd$ the energy of weakly decaying $B$ mesons,
corrected for detector acceptance and resolution; $\xel$ stands for
the corrected scaled energy of the leading $B$ meson, that is the
first meson produced in the fragmentation process, which can also be a
heavier resonance ($B^\star$, $B^{\star\star}$).

\begin{table}[htbp]
  \begin{center}
    \begin{tabular}{|c|lll|}
\hline
\rule{0pt}{4.4mm}
Channel & \multicolumn{3}{c|}{Decay chain} \\\hline \rule{0pt}{4.4mm}
1 & $B^0\to \ell \nu D^\star$ & $D^\star\to D^0 \pisoft$ & $ D^0\to K \pi$          \\ \rule{0pt}{4.4mm}
2 & $B^0\to \ell \nu D^\star$ & $D^\star\to D^0 \pisoft$ & $ D^0\to K \pi \pi \pi$  \\ \rule{0pt}{4.4mm}
3 & $B\to \ell \nu D^0      $ &                          & $ D^0\to K \pi$          \\ \rule{0pt}{4.4mm}
4 & $B^0\to \ell \nu D      $ &                          & $ D\to K \pi \pi$        \\ \rule{0pt}{4.4mm}
5 & $B^0\to \ell \nu D^\star$ & $D^\star\to D^0 \pisoft$ & $ D^0\to K \pi \pi^0$ \\\hline
    \end{tabular}
    \caption{\footnotesize $B$-decay channels used in the analysis.}
    \label{tab:channels}
  \end{center}
\end{table}

The analysis uses the full LEP I statistics collected by ALEPH between
1991 and 1995, amounting to almost four million hadronic $Z$ decays.
Recently this data set has been reprocessed using improved
reconstruction algorithms.  The main benefits for this analysis are
related to the enhanced secondary vertex reconstruction efficiency and
the improved particle identification.  A discussion of the
reprocessing can be found in~\cite{bs}.

After a description of the ALEPH detector, the selection of
$B\to \ell \nu D^{(\star)}$ decays is detailed in
Section~\ref{sec:selection}.  In Section~\ref{sec:erec} the
reconstruction of the $B$ meson energy is described, followed by the
extraction of the spectrum and comparison with the predictions of
different models in Section~\ref{sec:misura}.  Systematic errors are
discussed in Section~\ref{sec:syst}, and checks on the
self-consistency and robustness of the analysis are presented in
Section~\ref{sec:checks}.

\section{The ALEPH detector}
\label{sec:dete}
The ALEPH detector and its performance are described in detail
elsewhere~\cite{aleph,aleph2}. A high resolution vertex detector
(VDET) consisting of two layers of silicon with double-sided readout
measures $r \phi$ and $z$ coordinates at average radii of 6.5~cm and
11.3~cm, with $12~\mu$m resolution at normal incidence.  The VDET
provides full azimuthal coverage, and polar angle coverage to $|\cos
\theta|<0.85$ for the inner layer and $|\cos \theta|<0.69$ for both
layers. Outside VDET, particles traverse the inner tracking chamber
(ITC) and the time projection chamber (TPC).  The ITC is a cylindrical
drift chamber with eight axial wire layers with radii between 16 and
26~cm.  The TPC measures up to 21 space points per track at radii
between 40 and 171~cm, and also provides a measurement of the specific
ionization energy loss ($\dedx$) of each charged track. These three
detectors form the tracking system, which is immersed in a 1.5~T axial
magnetic field provided by a super-conducting solenoid. The combined
tracking system yields a transverse-momentum resolution of
$\sigma(p_T)/p_T = 6 \times 10^{-4} p_T\mathrm{(GeV/\mathit{c})}
\oplus 0.005$. The resolution of the three-dimensional impact
parameter for tracks having two VDET hits
can be parametrised as $\sigma = 25~\mu\mathrm{m}+ 95
\mu\mathrm{m}/p$, ($p$ in GeV/$c$).

The electromagnetic calorimeter (ECAL) is a lead/wire chamber sandwich
operated in proportional mode. The calorimeter is read out in
projective towers that subtend typically $0.9^\circ \times 0.9^\circ$,
segmented in three longitudinal sections. The hadron calorimeter
(HCAL) uses the iron return yoke as absorber. Hadronic showers are
sampled by 23 planes of streamer tubes, with analogue projective tower
and digital hit pattern readouts.  The HCAL is used in combination
with two double layers of muon chambers outside the magnet for muon
identification.

\boldmath
\section{Selection of $B\to \ell \nu D^{(\star)}$ decays}
\unboldmath
\label{sec:selection}
A Monte Carlo simulation based on JETSET 7.4~\cite{jetset} and tuned
to ALEPH data~\cite{jetsetaleph,megapapero} has been used in order to
extract resolution functions, acceptance corrections and background
compositions.  About five million $b\bar b$ events were simulated, and
more than twice the data statistics of $q \bar q$ events.  The present
analysis uses $b \bar b$ events to determine the $\xewd$ and $\xel$
spectrum starting from observed $\xerec$ spectra, and $q \bar q$
events to evaluate the \udsc\ component of the selected sample.

The decays $B\to \ell \nu D^{(\star)}$ are searched for in hadronic
events, containing at least one lepton (electron or muon) identified
using standard criteria~\cite{leptons}.  The momentum cut used to
define lepton candidates is $p>2$~GeV/$c$ for electrons and
$p>2.5$~GeV/$c$ for muons.  The transverse momentum $p_T$ of the
lepton with respect to the nearest jet, with the lepton excluded from
the jet, is required to be larger than 1~GeV/$c$, which helps
rejecting fake candidates and leptons not coming from direct decays of
$b$ hadrons. Both electron and muon candidates are required to have a
measured $\dedx$ compatible with the expected value.

Events are divided into two hemispheres by  a plane perpendicular to
the thrust axis; in each hemisphere containing a lepton a $D$ meson
reconstruction is attempted in the decay modes described in
Table~\ref{tab:channels}.  At least two charged tracks from the $D$
meson decay are required to have VDET hits, in order to ensure a good
reconstruction of the $D$ vertex position and to reject combinatorial
background.  Loose cuts are applied to track momenta, in order to
minimise the bias in the $B$ momentum distribution.  Tracks are not
considered as kaon candidates if their measured ionization is
incompatible with the kaon hypothesis by more than three standard
deviations. The charge of the kaon candidate is required to be the
same as that of the lepton, as expected for semileptonic $B$ meson
decays.
 
Tracks assigned to a $D$ meson decay are fitted to a common vertex,
and the track combination is rejected if the $\chi^2$
of the fit is larger than 20. If more than one combination fulfils
this requirement for channels 3 and 4, the one with the smallest
$\chi^2$ is chosen.  In channel 5, the $\pi^0$ closest in angle to the
charged pion is selected and added to form the $D^0$.

For channels 1, 2 and 5, a soft pion $\pisoft$ is added to the $D$
candidate to form a $D^{\star}$ meson; the $\pi_{\mathrm s}$ momentum
is required to be larger than 250~MeV/$c$ and smaller than 3~GeV/$c$.
The difference between the reconstructed $D^{\star}$ and $D$ masses is
required to be within 5~MeV/$c^2$ of its nominal value.  In the case
of multiple candidates in a given hemisphere, the track combination is
chosen for which the reconstructed $(D^\star-D)$ mass difference is
closest to the nominal value.

A vertex fit is performed using the $D$ candidate and the lepton
track, and again the combination of tracks is rejected if the
$\chi^2$ of the fit is larger than 20.  The $B$
vertex is required to lie between the interaction point, reconstructed
event-by-event, and the $D$ vertex.

Channels 3 and 4 are further enriched in signal events using harder
 cuts on the kaon; in addition a $\pi_{\mathrm s}$ veto is
applied: if a track is found which is compatible with the
reconstructed $B$ vertex and the combination track-$D$ candidate has a
mass close to the mass of the $D^\star$, the candidate is discarded.
This procedure reduces the overlap between channels at the permil level.
Finally, tighter cuts on the reconstructed $D$ mass and the $\chi^2$
of the vertex fit are imposed.

\begin{figure}[tbp]
  \begin{center}
    \epsfig{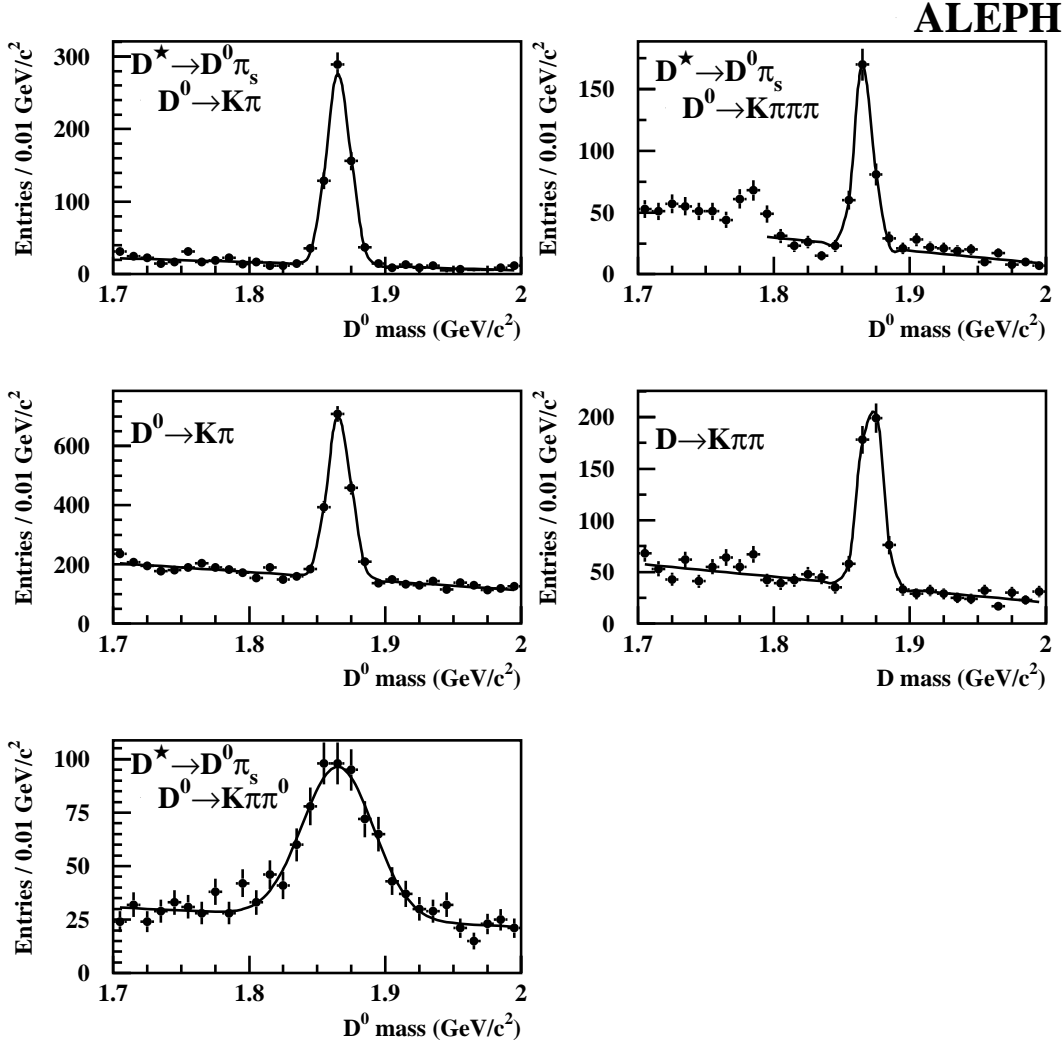}
    \caption{\footnotesize Reconstructed $D$ mass peaks in the five channels. In the second channel the second peak at lower mass comes from the decay channel $D^0\to K \pi\pi\pi \pi^0$ and is excluded from the fit.}
    \label{fig:d0mass}
  \end{center}
\end{figure}

The $D$ mass spectra are shown in Fig.~\ref{fig:d0mass}.  The
reconstructed $D$ mass peaks are fitted in a region between 1.7 and
2.0~$\mathrm{GeV}/c^2$ with a Gaussian and a linear component.
Table~\ref{tab:evsel} shows the chosen $D$ mass windows, the number of
reconstructed candidates and the fitted Gaussian fractions.

 \begin{table}[b]
   \begin{center}
     \begin{tabular}{|ll|c|c|c|c|}
 \hline
 \multicolumn{2}{|c|}{\rule{0pt}{4.4mm}
 Channel}                       & $D$ window   & Events & Resolution   & Gaussian \\
                           &                        & (MeV/$c^2$)  &        &  (MeV/$c^2$) &     Fraction    \\\hline     
 \rule{0pt}{4.4mm}
 $D^\star \to D^0 \pisoft$ & $\!\!\! D^0\to K \pi $        & $1864\pm 30$ &   665  &    $8.3$     &  89 \% \\
 \rule{0pt}{4.4mm}
 $D^\star \to D^0 \pisoft$ & $\!\!\! D^0\to K \pi \pi \pi$ & $1864\pm 30$ &   388  &    $6.2$     &  69 \% \\
 \rule{0pt}{4.4mm}
                           & $\!\!\! D^0\to K \pi$         & $1864\pm 15$ &   1079  &    $8.4$     &  81 \% \\
 \rule{0pt}{4.4mm}
                           & $\!\!\! D\to K \pi \pi$       & $1869\pm 30$ &   580  &    $7.4$     &  64 \% \\
 \rule{0pt}{4.4mm}
 $D^\star \to D^0 \pisoft$ & $\!\!\! D^0\to K \pi \pi^0 $  & $1864\pm 50$ &   693  &    $25$      &  63 \% \\\hline
     \end{tabular}
     \caption{\footnotesize For the five channels, the $D$ mass window, the number of events in the window, the mass resolution and the Gaussian fraction.}
     \label{tab:evsel}
   \end{center}
 \end{table}

The fractions of the Gaussian components measured in the Monte Carlo
are compatible with those in the data within statistical errors, while the
widths are about $5-10\%$ smaller; this is taken into account in the
evaluation of the systematic uncertainties.

\boldmath
\section{$B$ energy reconstruction}
\unboldmath
\label{sec:erec}
The scaled energy of the weakly-decaying $B\to \ell \nu D^{(\star)}$ hadron is estimated as
\begin{equation}
  \label{eq:xeb}
  \xerec = \frac{E_\ell + E_{D^{(\star)}}+E_\nu}{E_{\mathrm{beam}}} \quad .
\end{equation}
The terms $E_{D^{(\star)}}$ and $E_\ell$ are provided by the
direct reconstruction, while the neutrino energy $E_\nu$
is estimated from the missing energy in the hemisphere:
\begin{equation}
  \label{eq:nue}
  E_\nu = E^{\mathrm{hemi}}_{\mathrm{tot}} - E^{\mathrm{hemi}}_{\mathrm{vis}} \quad ,
\end{equation}
where $E^{\mathrm{hemi}}_{\mathrm{tot}}$ is estimated taking into account the measured mass in both hemispheres~\cite{neutrinores}:
\begin{equation}
  \label{eq:etot}
  E^{\mathrm{hemi}}_{\mathrm{tot}} = E_{\mathrm{beam}} + \frac{m^2_{\mathrm{same}}-m^2_{\mathrm{oppo}}}{4E_{\mathrm{beam}}} \quad .
\end{equation}
Both charged and neutral particles are used in Eqns.~(\ref{eq:nue})
and~(\ref{eq:etot}).  In the lepton hemisphere neutral hadronic energy
is expected to come only from fragmentation. Therefore, in order to
avoid spurious calorimetric fluctuations, its contribution is taken
into account only outside a cone of 10 degrees of half opening angle around each of the B
meson decay products.
Table~\ref{tab:resolu} shows the resolution on $\xeb$
estimated on simulated $b\bar b$ events; the distributions
are well described by two Gaussians, accounting for core and tails.

\begin{table}[htbp]
  \begin{center}
    \begin{tabular}{|ll|c|c|c|}
\hline
 \multicolumn{2}{|c|}{\rule{0pt}{4.4mm}
Channel}                    &  Core (\%)   &  Core  &  Tail   \\
                          &             &          &    resolution   &resolution      \\ \hline \rule{0pt}{4.4mm}
$D^\star \to D^0 \pisoft$ & $D^0\to K \pi $       & 57  &  0.03  &   0.10 \\ \rule{0pt}{4.4mm}
$D^\star \to D^0 \pisoft$ & $D^0\to K \pi \pi \pi$&  54  &  0.04  &  0.12 \\ \rule{0pt}{4.4mm}
                          & $D^0\to K \pi$        &  61  &  0.05  &   0.15 \\ \rule{0pt}{4.4mm}
                          & $D\to K \pi \pi$     &  65  &  0.05  &   0.15 \\ \rule{0pt}{4.4mm}
$D^\star \to D^0 \pisoft$ & $D^0\to K \pi \pi^0 $ &  57  &  0.04  &  0.11 \\\hline
    \end{tabular}
    \caption{\footnotesize For the five channels the  $\xeb$ resolution on simulated events. The resolution can be parametrised with two Gaussians, describing the core and the tails.}
    \label{tab:resolu}
  \end{center}
\end{table}
\section{Unfolding methods}
\label{sec:misura}
\begin{figure}[htbp]
  \begin{center}
    \epsfig{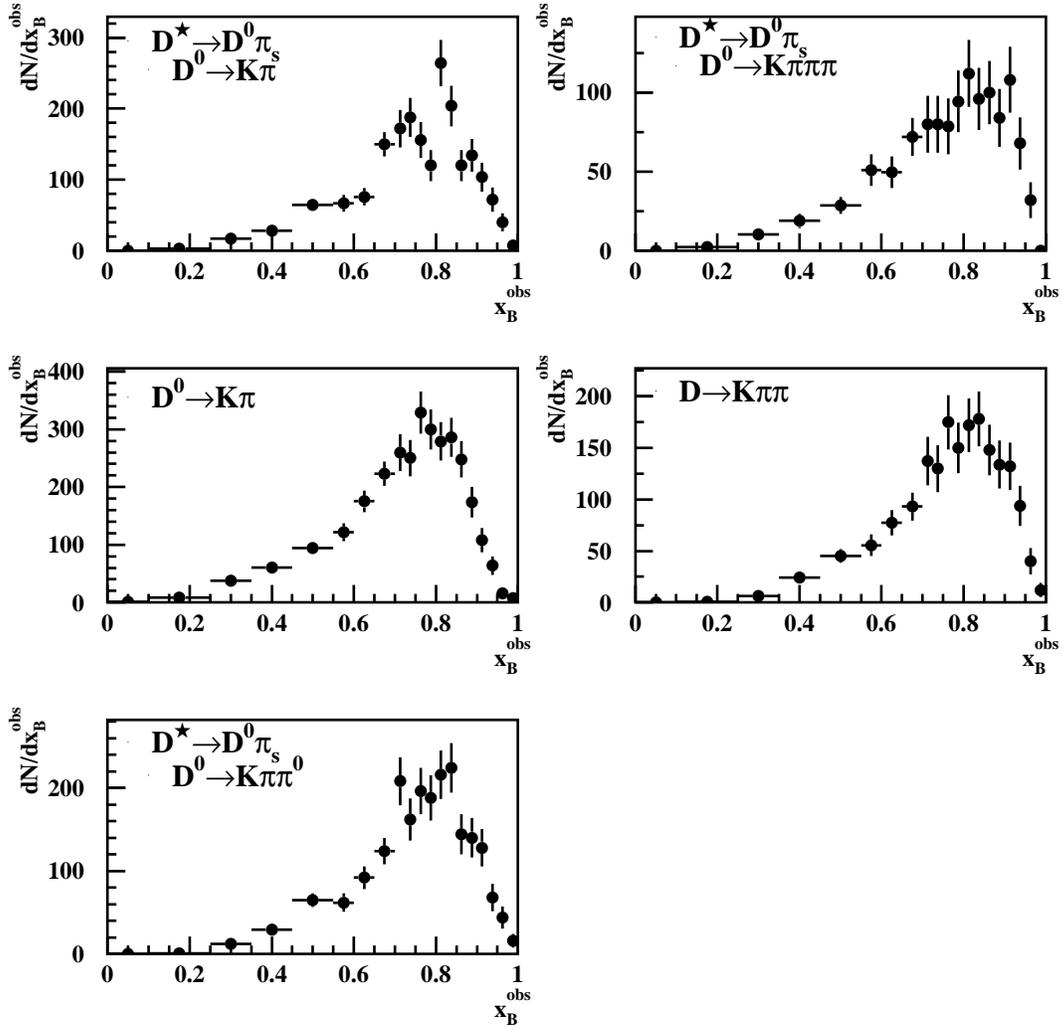}
    \caption{\footnotesize In the five channels, $\xerec$ spectra, after \udsc\ background subtraction
      and before acceptance corrections.}
    \label{fig:spectra}
  \end{center}
\end{figure}

The scaled energy of the weakly-decaying $B$ mesons, $\xerec$, is
reconstructed in 20 bins between 0 and 1 with a variable width.  In
each bin, the \udsc\ background is estimated using the simulation, and
subtracted from the spectrum. This amounts to about 2\% of the events,
concentrated mostly at low $\xerec$.  The measured spectra after
subtraction are shown in Fig.~\ref{fig:spectra}.

With these events two different kinds of analyses can be performed:
\begin{itemize}
\item a model-dependent analysis, in which different fragmentation models 
available in the literature are tuned to fit the observed spectra;
\item a model-independent analysis, in which the shapes of $\xewd$ and $\xel$ 
are reconstructed by correcting the observed spectra for detector acceptance, 
 resolution and missing particles.
\end{itemize}

\subsection{Model-dependent analysis}
\label{subsec:moddep}
Various fragmentation functions $D^H_b(z)$ are implemented in the
Monte Carlo generator JETSET 7.4, which also simulates initial and
final state photon radiation and hard gluon emission. The
reconstructed spectra obtained from the simulated $b\bar b$ events are
tuned to best reproduce the $\xerec$ distribution observed in the
data, by minimising the global $\chi^2$.  The following
parametrisations for $D^H_b(z)$ are used:
\begin{eqnarray}
\nonumber   \mathrm{\small Peterson~\textit{et \textrm{al.}}\ \cite{peterson}}:&D^H_b(z) \propto & \frac{1}{z} \left( 1- \frac{1}{z}-\frac{\epsilon_b}{1-z}\right)^{-2} \quad , \\
\nonumber     \mathrm{\small Kartvelishvili~\textit{et~\textrm{al.}}\ \cite{kart}}:&D^H_b(z) \propto & z^{\alpha_b} \left(1-z\right)
\quad , \\
\nonumber     \mathrm{\small Collins~\textit{et~\textrm{al.}}\ \cite{collins}}:&D^H_b(z) \propto & \left( \frac{1-z}{z}+\frac{\left( 2-z\right) \epsilon_b}{1-z}\right)\times\\
\nonumber &&  \left( 1+z^2\right) \left( 1- \frac{1}{z}-\frac{\epsilon_b}{1-z}\right)^{-2} \quad .
  \label{eq:models}
\end{eqnarray}

The minimisation is performed with respect to the free parameter of
each model, and the $\chi^2$ is written as:

\begin{equation}
 \chi^2 \,= \,\sum_{c=1}^{5} \,\sum_{i=1}^{20} \, \frac{ \left[ n^{\mathrm{DT}}_i (c)- n^{\mathrm{MC}}_i (c)\right]^2 }
{  \left[ \sigma^{\mathrm{DT}}_i (c)  \right]^2 + \left[ \sigma^{\mathrm{MC}}_i (c)\right]^2   }  \quad ,
\label{eq:chifit}
\end{equation}
where $c$ runs over the channels used, $i$ runs over the $\xerec$ bins
defined as in Table~\ref{tab:tuttoxeb}, $n^{\mathrm{DT}}$ and
$n^{\mathrm{MC}}$ are the number of candidates per channel and per bin
observed in the data and expected from the Monte Carlo, normalised to
the same number of entries. The quantities $\sigma$ are defined as statistical uncertainties.

Table~\ref{tab:models} shows the fitted values for the different model parameters, 
together with statistical and  systematic uncertainties  from the sources 
discussed in Section~\ref{sec:syst}. Also shown are the values for the
mean scaled energy.

\begin{table}[htbp]
  \begin{center}
    \begin{tabular}{|l||c|c|}
      \hline
      Model & Fit results & Mean energies \\\hline\hline
      Peterson & $\epsilon_b = \left(31\pm 3 \pm 5\right)\! \times\!  10^{-4}$ & $\mxewd=\left(700\pm 4 \pm 5\right)\! \times \! 10^{-3}$ \\
    \cite{peterson}   &   $\chi^2/N_{\mathrm{DOF}}=117/94$ & $\mxel=\left(721\pm 4 \pm 5\right)\!  \times\!  10^{-3}$\\\hline
      Kartvelishvili &  $\alpha_\beta=13.7\pm 0.7 \pm 1.1$ & $\mxewd=\left(713\pm 4 \pm 6\right)\!  \times\!  10^{-3}$ \\
    \cite{kart}     &   $\chi^2/N_{\mathrm{DOF}}=107/94$ & $\mxel=\left(734\pm 4 \pm 6\right)\!  \times\!  10^{-3}$\\\hline
      Collins & $\epsilon_b =\left(185\pm 25 \pm 41\right)\! \times\!  10^{-5}$ &  $\mxewd=\left(681\pm 4 \pm 5\right)\!  \times\!  10^{-3}$ \\
    \cite{collins}  &   $\chi^2/N_{\mathrm{DOF}}=181/94$ & $\mxel=\left(701\pm 4 \pm 5\right)\!  \times\!  10^{-3}$\\\hline
    \end{tabular}
    \caption{\footnotesize Fit results with different fragmentation models.
 The systematic errors account for the sources of uncertainties discussed in Section~\ref{sec:syst}.
 The $\chi^2/N_{\mathrm{DOF}}$ is calculated using statistical errors only.}
     \label{tab:models}
   \end{center}
 \end{table}
The Kartvelishvili model describes the data slightly better than the
Peterson model. The Collins model is clearly disfavoured.

\subsection{Model-independent analysis}
\label{sec:boh}
The $\xewd$ and $\xel$ spectra are obtained by correcting the observed 
$\xerec$ spectra for acceptance, detector resolution and missing particles.

The normalised binned spectrum $ f_i\!  \left(\xewd \right)$ can be obtained using the relation
\begin{equation}
  \label{eq:spettrovero}
  f_i\! \left(\xewd\right) \,= \, \frac{1}{T} \, \sum_{c=1}^{5} \, 
\frac{1}{\epsilon_i^{\mathrm {wd}}{(c)}} \, 
\sum_{j=1}^{20} \, G_{ij}^{\mathrm {wd}}(c) \, n^{\mathrm{DT}}_j (c) \quad ,
\label{eq:gij}
\end{equation}
where  $\epsilon_i^{\mathrm {wd}}(c)$ is the acceptance correction in bin $i$ for channel $c$; 
$n^{\mathrm{DT}}_j (c)$ is the number of reconstructed $B$ mesons in the data for channel $c$, 
with a measured energy falling in bin $j$; $G_{ij}^{\mathrm {wd}}(c)$ is the resolution 
matrix that links mesons with $\xerec$ in bin $j$ and $\xewd$ in bin $i$, for channel $c$; 
$T$ is the normalisation factor defined by the condition  $\sum_i f_i =1$. A similar equation holds
for the extraction of  $f_i\! \left(\xel\right)$, where the effect of the missing particles from
$B^\star$ and $B^{\star\star}$ decays is folded in $\epsilon^{\mathrm {L}}$ and $G^{\mathrm {L}}$.

The acceptance corrections $\epsilon_i$ and the resolution matrix
$G_{ij}$ are taken from the simulation. 
\begin{figure}[!t]
  \begin{center}
    \epsfig{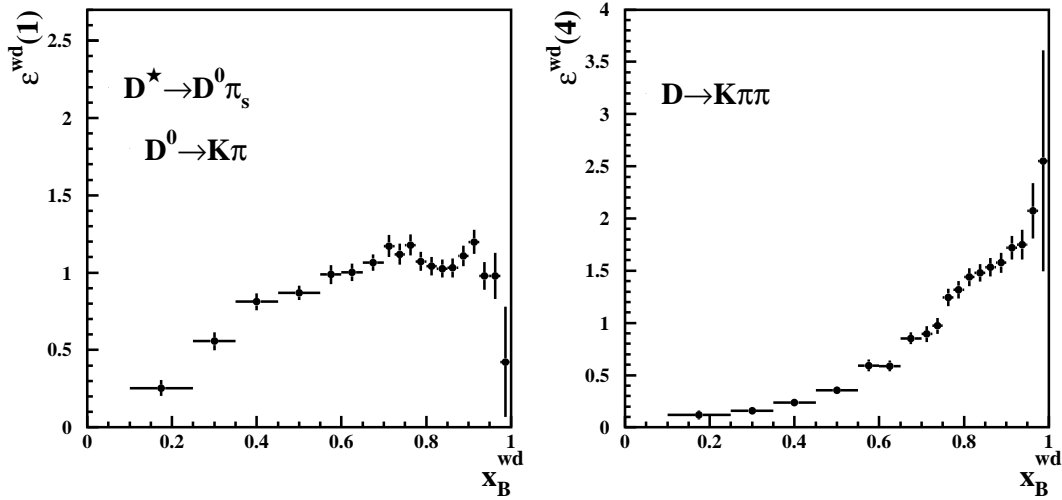}
    \caption{\footnotesize Acceptance corrections  $\epsilon_i^{\mathrm {wd}}(c)$ for $\xewd$. 
      The absolute scale is chosen as to conserve the total number of
      selected events for each channel. Only the distributions for channels
      1 and 4 are shown, since they represent the extreme
      behaviours.}
    \label{fig:acceptances}
  \end{center}
\end{figure}
The acceptance corrections show a different behaviour among the
different channels; the two extreme situations are shown in
Fig.~\ref{fig:acceptances}.  A dependence on the fragmentation
function present in Monte Carlo is induced in the measured spectrum
through $G_{ij}$; hence, the Monte Carlo used to calculate $G_{ij}$
must be reweighted to the best estimate of $f_i\left(\xewd\right)$ from
data.  This is done using an iterative procedure, calculating $f^N_i\!
\left(\xewd\right)$ using the $G^{N-1}_{ij}$ from Monte Carlo
reweighted to $f^{N-1}_i\!  \left(\xewd\right)$.  The weights
\mbox{$w_i\equiv {f^{N-1}_i(\mathrm{DT})}/{f_i(\mathrm{MC})}$} are applied to standard
Monte Carlo events; to avoid fluctuations due to the limited
statistics in data events, the distribution of the weights $w_i$ is
smoothed with a polynomial function.  Possible systematic effects
related to the smoothing are studied in Section~\ref{sec:checks}.

The whole procedure is then repeated until the change in $f_i$ in consecutive iterations is a small fraction of the statistical errors.

The statistical error matrix $E_{ij}$ is  calculated by repeating $20\times 5$
analyses, varying in each of them the quantities  $n^{\mathrm{DT}}_j
 (c)$ by one standard deviation:
\begin{equation}
  E_{ij} = \sum_{c=1,5} \sum_{k=1,20} (f_i^{(ck)}-f_i^{{\rm STD}})(f_j^{(ck)}-f_j^{{\rm STD}}) \quad ,
\end{equation}
where $f_i^{(ck)}$ is the result of the convergence for $f_i$ when
$n^{\mathrm{DT}}_k (c)$ is varied by its statistical error, and
$f_i^{{\rm STD}}$ the nominal result.

The results, together with the statistical and systematic errors, are
given in Table~\ref{tab:tuttoxeb}. The full error matrices are shown
in the Appendix.

From the binned spectra $f_i\! \left(\xewd\right)$ the mean value is
calculated as
\begin{equation}
  \label{eq:mxel}
  \mxewd = \sum_{i=1}^{20} x_i \;  f_i\! \left(\xewd\right) \quad ,
\end{equation}
where $x_i$ is the central value of bin $i$.  The bin size chosen is
such that the deviation from linearity of the distribution within a bin is
negligible.

The statistical error on $\mxeb$ is calculated using the same procedure used as for $f_i$.

The results for $\mxewd$ and $\mxel$ are
\begin{eqnarray*}
\mxewd &=& 0.7163 \pm 0.0061  \, (\mathrm{stat}) \quad , \\
\mxel  &=& 0.7361 \pm 0.0063  \, (\mathrm{stat}) \quad .
\end{eqnarray*}

\section{Systematic errors}
\label{sec:syst}
Possible systematic effects due to uncertainties on the physics
parameters used in the Monte Carlo, limited accuracy in the simulation
of the detector response, or effects intrinsic to the analysis method have
been investigated.

The physics parameters used in the Monte Carlo simulation that are
relevant for the analysis are adjusted to the most recent experimental
measurements and varied within their estimated uncertainty by
reweighting simulated events.  The effect of the reweighting
propagates to the results through the resolution matrix $G_{ij}(c)$
and the acceptance corrections $\epsilon_i(c)$, which are taken from
the simulation.  The differences from the standard results are taken
as systematic errors.

The sources of uncertainty considered are:

\begin{table}[tbp]
  \begin{center}
    \begin{tabular}{|l|c|}
      \hline
 \rule{0pt}{4.4mm}
       Process                              &   BR(\%)     \\\hline
 \rule{0pt}{4.4mm}
       $B\to D \ell \nu$                    & $1.95\pm 0.27$\\
 \rule{0pt}{4.4mm}
       $B\to D^\star \ell \nu$              & $5.05\pm 0.25$ \\
 \rule{0pt}{4.4mm}
       $B\to D^{(\star)} X \ell \nu$          & $2.7\pm 0.7$   \\
 \rule{0pt}{4.4mm}
       \quad with $B\to D_1 \ell \nu$       & $0.63\pm 0.11$ \\
 \rule{0pt}{4.4mm}
       \quad with $B\to D^\star_2 \ell \nu$ & $0.23\pm 0.09$ \\
 \rule{0pt}{4.4mm}
       $b\to u \ell \nu$                    & $0.15\pm 0.10$ \\\hline
 \rule{0pt}{4.4mm}
       $\sum B\to \ell \nu X$               & $9.85\pm 0.80$ \\\hline
 \rule{0pt}{4.4mm}
       Inclusive $B\to \ell \nu X$          &$10.18\pm 0.39$ \\\hline

    \end{tabular}
    \caption{\footnotesize Exclusive branching ratios for the  $B\to \ell \nu X$ process~\cite{pdg,alephsld}. 
      The sum is consistent with the measurement of the inclusive
      $B\to \ell \nu X$ rate.}
    \label{tab:slbr2}
  \end{center}
\end{table}

\begin{itemize}
\item Semileptonic decays of $B$ mesons. 
  
  The current experimental knowledge~\cite{pdg,alephsld} of the
  semileptonic branching ratios of $B$ mesons is summarised in
  Table~\ref{tab:slbr2}. The sum of the exclusive (or semi-exclusive)
  rates is consistent within errors with the inclusive measurement of
  \mbox{$\mathrm{BR}(B\to \ell \nu X)$}.
  
  The analysis is not sensitive to the total $\mathrm{BR}(B\to \ell \nu X)$, but is affected by a change in the relative rates of the
  different components, since these contribute in a different way to the
  average acceptance corrections and resolution matrix.

  Six sources of systematic error are  calculated using the values in Table~\ref{tab:slbr2}:
  \begin{enumerate}
  \item The inclusive \mbox{$\mathrm{BR}(B\to D^{(\star)}X  \ell \nu)$} is varied
    within its experimental error.
    \begin{displaymath}
      \nonumber   \Delta \mxewd  = 0.0019 \quad \Delta \mxel = 0.0020
    \end{displaymath}
  \item The rate for the narrow $D_1$ state is varied by its
    experimental error, while leaving the total
    \mbox{$\mathrm{BR}(B\to D^{(\star)} X \ell \nu)$} at its
    central value.
    \begin{displaymath}
      \nonumber   \Delta \mxewd = 0.0001 \quad  \Delta \mxel = 0.0001
    \end{displaymath}
  \item The rate for the narrow $D^\star_2$ state is varied by
    its experimental error, while leaving the total
    \mbox{$\mathrm{BR}(B\to D^{(\star)} X \ell \nu)$} at its
    central value.
    \begin{displaymath}
      \nonumber  \Delta \mxewd = 0.0001 \quad  \Delta \mxel = 0.0001
    \end{displaymath}
  \item The rate of wide $D^{\star \star}$ states, not yet measured, is put to
    zero and compensated with non-resonant $D^{(\star)} \ell \nu \pi$
    final states, thus leaving the total \mbox{$\mathrm{BR}(B\to
      D^{(\star)} X \ell \nu)$} at its central value.
    \begin{displaymath}
      \nonumber  \Delta \mxewd  = 0.0017 \quad \Delta \mxel = 0.0016
    \end{displaymath}
  \item The $\mathrm{BR}(B\to D \ell \nu )$ is varied by its
    experimental error:
    \begin{displaymath}
      \nonumber   \Delta \mxewd = 0.0008 \quad \Delta \mxel = 0.0007
    \end{displaymath}
  \item The $\mathrm{BR}(B\to D^\star \ell \nu)$ is varied by its
    experimental error:
    \begin{displaymath}
      \nonumber  \Delta \mxewd  = 0.0008 \quad \Delta \mxel = 0.0008
    \end{displaymath}

  \end{enumerate}
  
\item Missing particles from $B^{\star\star}$ production. 
  
  When deriving the energy spectrum
  of the leading $B$ meson, the correction due the energy carried away by the pion produced 
  in the $B^{\star\star}$ decay enters in the resolution matrix and the acceptance corrections. 
  The rate of  $b\to B^{\star\star}$ is varied within its experimental error: 
  $f_{B^{\star\star}} = 0.299 \pm 0.058$~\cite{cdf,boh,boh2}, and the resulting systematic error 
  is $\Delta \mxel = 0.0025$. The weakly-decaying $B$ meson spectrum
  is not affected by this source of uncertainty.
\item Modelling of the $B^{\star\star}$ production. 
  
  From spin counting, the relative rates
  of ($B_1$, $B^{\star}_0$, $B^{\star}_1$, $B^{\star}_2$) are predicted to be (3,1,3,5)~\cite{bstar}; 
  changing this to (1,1,1,1) gives a systematic error of $\Delta \mxel = 0.0004$.
\item Production of $B^{\star}$ from $b$ quarks.
  
  Due to the small mass difference between $B^{\star}$ and $B$ mesons, the effect of $B^{\star}$
  production in $b$ quark fragmentation is found to be much smaller than for $B^{\star\star}$,
  and it is completely negligible for the present analysis.
  
\end{itemize}

The relevant sources of systematic uncertainties due to the detector simulation 
are identified to be:
\begin{itemize}
\item Neutrino energy reconstruction.  
  
  The accuracy of the neutrino energy reconstruction is checked in
  had\-ronic events enriched in light primary quarks and in had\-ronic
  decays of $\tau \tau$ events.  In the first sample, a ``fake''
  neutrino is simulated by removing a charged particle from the
  reconstructed event; its energy is then reconstructed using
  Eqns.~(\ref{eq:nue}) and (\ref{eq:etot}).  The method can be applied
  both to data and Monte Carlo events, determining the bias between the
  reconstructed energy and the momentum measured with the tracking
  system. Such a bias is found to be reproduced by the Monte Carlo
  with a precision better than 50~MeV, for all momenta of the
  deleted track. In the second sample, where the event topology is much
  simpler, the energy of the ``reconstructed'' $\nu_\tau$ is compared
  between data and Monte Carlo events. Also in this case the worst
  discrepancy observed is smaller than 50~MeV.  This value is used as
  a conservative estimate of the systematic uncertainty on the
  neutrino energy, resulting in $\Delta \mxewd = 0.0023$ and $\Delta
  \mxel = 0.0023$.

\item Vertexing and charm meson reconstruction.
  
  If the purity and the kinematic properties of the selected
  candidates are not well described by the simulation, the acceptance
  corrections and resolution matrices can be inadequate. In order to
  check for these effects, the distributions of the $\chi^2$
  probability for the reconstructed $D$ vertices are compared, channel
  by channel, with the simulation. Small differences are observed, and
  the Monte Carlo distribution is reweighted in order to reproduce the data.
  The shift in the corrected average energy is taken as systematic
  uncertainty. The resulting error estimates are $\Delta \mxewd =
  0.0001$ and $\Delta \mxel = 0.0001$.
  
  Furthermore, the reconstructed $D$ mass distributions in data and
  Monte Carlo are compared.  In simulated events the widths of the
  mass spectra are found to be $5-10\%$ smaller, while the fractions
  of the Gaussian components, estimated from a fit to the sidebands,
  are reproduced within their statistical error of about 5\%. The mass
  cuts reported in Table~\ref{tab:evsel} are adjusted in order to take into
  account both effects, taking the total shift in the extracted energy
  spectrum as systematic uncertainty.  The resulting estimates are
  \begin{eqnarray}
    \nonumber    \left.\Delta \mxewd\right|_{\textrm{\footnotesize{width}}}  = 0.0011 &\quad & \left.\Delta \mxel\right|_{\textrm{\footnotesize{width}}} = 0.0010 \quad ,\\
    \nonumber   \left.\Delta \mxewd\right|_{\textrm{\footnotesize{purity}}}= 0.0021  &\quad & \left.\Delta \mxel\right|_{\textrm{\footnotesize{purity}}} = 0.0021 \quad .
  \end{eqnarray}

\end{itemize}

Possible systematic effects related to the analysis procedure are:

\begin{itemize}
\item Background subtraction.
  
  As previously explained, a bin-by-bin subtraction of candidates not
  coming from $b \bar b$ events is performed before deriving the $B$
  meson energy spectra.  The efficiency for this kind of background
  has been extracted directly from data events, which have been
  enriched in background events by selecting wrong sign candidates. It
  is found to be compatible with Monte Carlo simulation within the
  statistical error of about 25\%. The background is varied within
  this range and the systematic errors associated are $\Delta \mxewd =
  0.0021$ and $\Delta \mxel = 0.0022$.

\item Monte Carlo statistics. 
  
  Statistics of simulated events are larger than for data events by a
  factor of 5. In order to evaluate the related uncertainty, the
  acceptance corrections $\epsilon_i(c)$ and the matrix elements
  $G_{ij}(c)$ are varied randomly by their statistical error in a
  series of toy experiments. The scatter of the results for $\mxewd$
  and $\mxel$ is taken as an estimate of the uncertainty due to the
  limited Monte Carlo statistics, yielding $\Delta \mxewd = 0.0029$
  and $\Delta \mxel = 0.0031$.
\end{itemize}

Adding in quadrature all the systematic contributions, the final results 
are:
\begin{eqnarray*}
\mxewd &=& 0.7163 \; \pm 0.0061  \, (\mathrm{stat}) \; \pm 0.0056 \, (\mathrm{syst}) \quad , \\
\mxel  &=& 0.7361 \; \pm 0.0063  \, (\mathrm{stat}) \; \pm 0.0063 \, (\mathrm{syst}) \quad . 
\end{eqnarray*}

The  bin-by-bin results for the measured spectra, with the
total systematic uncertainties, are shown in 
Table~\ref{tab:tuttoxeb}, while the statistical and total error matrices
are reported in the Appendix. The spectra are also shown in Fig.~\ref{fig:xewd}, where they are compared with the Monte Carlo predictions 
from different fragmentation models, with the free parameters fitted to
the data.

\begin{figure}[p]
  \begin{center}
    \epsfig{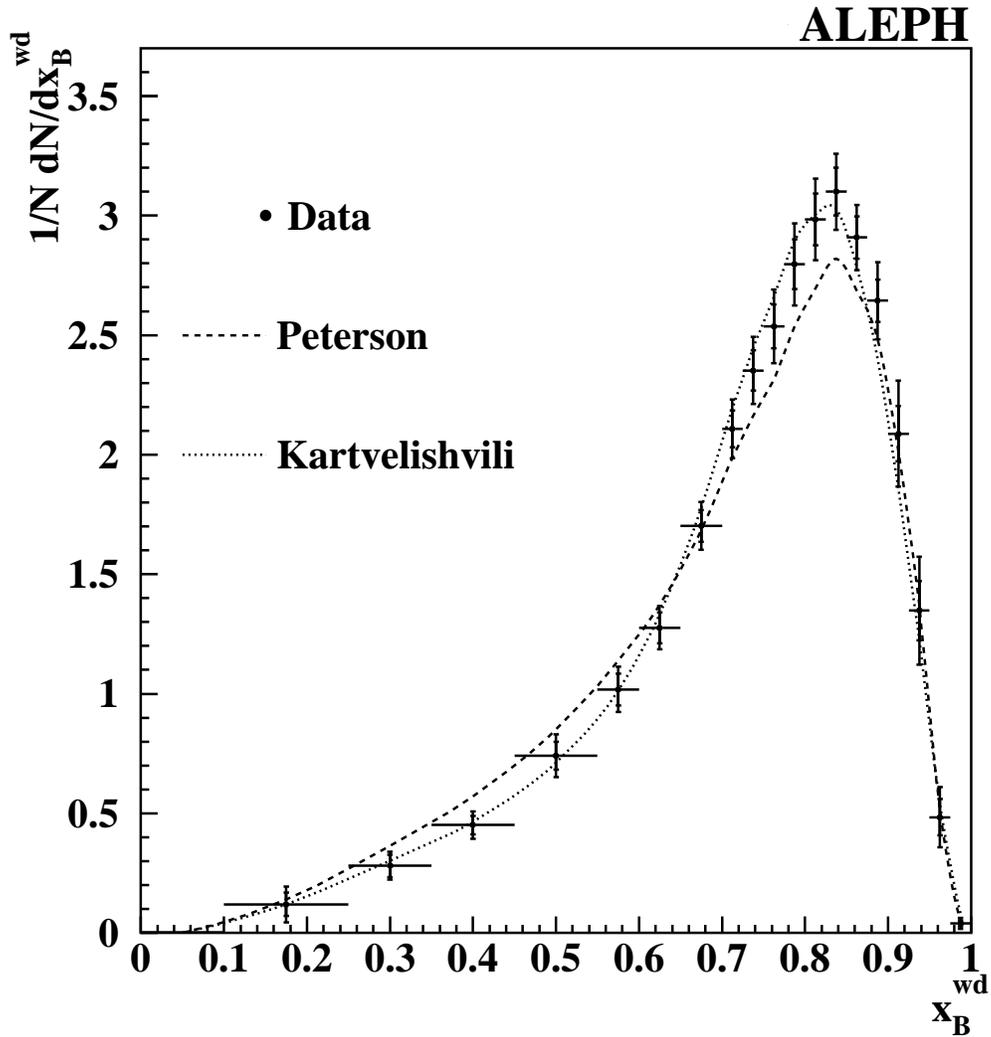}
    \caption{\footnotesize Scaled energy of the weakly-decaying $B$ meson, as reconstructed from data. 
      The inner error bars represent statistical errors, the larger
      ones the total uncertainties.  The best-fit distributions for
      the Peterson model and the Kartvelishvili model are
      superimposed. For the data, the bin-to-bin errors are highly
      correlated, as shown in the error matrices in the Appendix.}
    \label{fig:only2}
  \end{center}
\end{figure}

The models of Peterson and Kartvelishvili give the best agreement with
the data, and are compared with the $\xewd$ measurement in
Fig.~\ref{fig:only2}.

\begin{table}[ht!]
  \begin{center}
    \begin{tabular}{|r|l|r|r|}
      \hline 
      
      \rule{0pt}{4.8mm}
      Bin & $\xeb$ Range & \multicolumn{1}{c|}{$f_i\left(\xewd\right)$} & \multicolumn{1}{c|}{$f_i\left(\xel\right)$} \\
      [0.15cm]
      \hline
      1   & $0.\ \ \ \, \, \, -0.1  $    &  --$\;\;\;\;\;\;\;\;\;\;\;$  &  --$\;\;\;\;\;\;\;\;\;\;\;$   \\
      2   & $0.1\ \  \, \,    -0.25 $      &$17.9\pm 7.3\pm 8.6    $   &$17.6\pm 7.0\pm 6.4    $   \\
      3   & $0.25\ \, \,       -0.35$      &$28.1\pm 4.7\pm 3.5    $   &$26.4\pm 4.6\pm 3.6    $   \\
      4   & $0.35\ \,         -0.45 $      &$45.1\pm 3.9\pm 4.3    $   &$41.6\pm 3.8\pm 2.9    $   \\
      5   & $0.45\ \,         -0.55 $      &$74.1\pm 5.9\pm 6.6    $   &$65.1\pm 5.5\pm 5.7    $   \\
      6   & $0.55\ \,         -0.6  $      &$50.9\pm 3.3\pm 3.4    $   &$46.2\pm 3.1\pm 3.3    $   \\
      7   & $0.6\ \ \, \,     -0.65 $      &$63.8\pm 3.2\pm 3.2    $   &$55.8\pm 2.9\pm 3.0    $   \\
      8   & $0.65\ \,         -0.7  $      &$85.1\pm 3.3\pm 3.7    $   &$72.4\pm 2.9\pm 3.6    $   \\
      9   & $0.7\ \ \, \,     -0.725$      &$52.7\pm 1.9\pm 2.4    $   &$44.5\pm 1.7\pm 2.4    $   \\
      10  & $0.725            -0.75 $      &$58.8\pm 2.1\pm 2.8    $   &$50.3\pm 1.9\pm 2.9    $   \\
      11  & $0.75\ \,         -0.775$      &$63.4\pm 2.3\pm 3.1    $   &$54.8\pm 2.1\pm 3.0    $   \\
      12  & $0.775            -0.8  $      &$69.9\pm 2.6\pm 3.4    $   &$63.1\pm 2.5\pm 3.3    $   \\
      13  & $0.8\ \ \, \,     -0.825$      &$74.6\pm 2.7\pm 3.3    $   &$71.3\pm 2.7\pm 3.3    $   \\
      14  & $0.825            -0.85 $      &$77.5\pm 2.5\pm 3.1    $   &$76.5\pm 2.6\pm 2.7    $   \\
      15  & $0.85\ \,         -0.875$      &$72.7\pm 2.2\pm 2.6    $   &$81.0\pm 2.3\pm 2.2    $   \\
      16  & $0.875            -0.9  $      &$66.1\pm 2.2\pm 3.4    $   &$79.7\pm 2.1\pm 3.1    $   \\
      17  & $0.9\ \ \, \,     -0.925$      &$52.2\pm 2.9\pm 4.7    $   &$73.3\pm 3.0\pm 5.5    $   \\
      18  & $0.925            -0.95 $      &$33.7\pm 3.1\pm 4.7    $   &$53.8\pm 3.9\pm 6.6    $   \\
      19  & $0.95\ \,         -0.975$      &$12.1\pm 1.9\pm 2.5    $   &$23.7\pm 2.8\pm 4.3    $   \\
      20  & $0.975            -1.   $      &$ 1.0\pm 0.3\pm 0.5    $   &$ 2.7\pm 0.6\pm 0.8    $  \\\hline
    \end{tabular}
    \caption{\footnotesize The extracted spectra for the weakly-decaying $B$ meson and the leading $B$ meson. All numbers are given in units of $10^{-3}$.}
    \label{tab:tuttoxeb}
  \end{center}
\end{table}

\begin{figure}[p]
  \begin{center}
    \epsfig{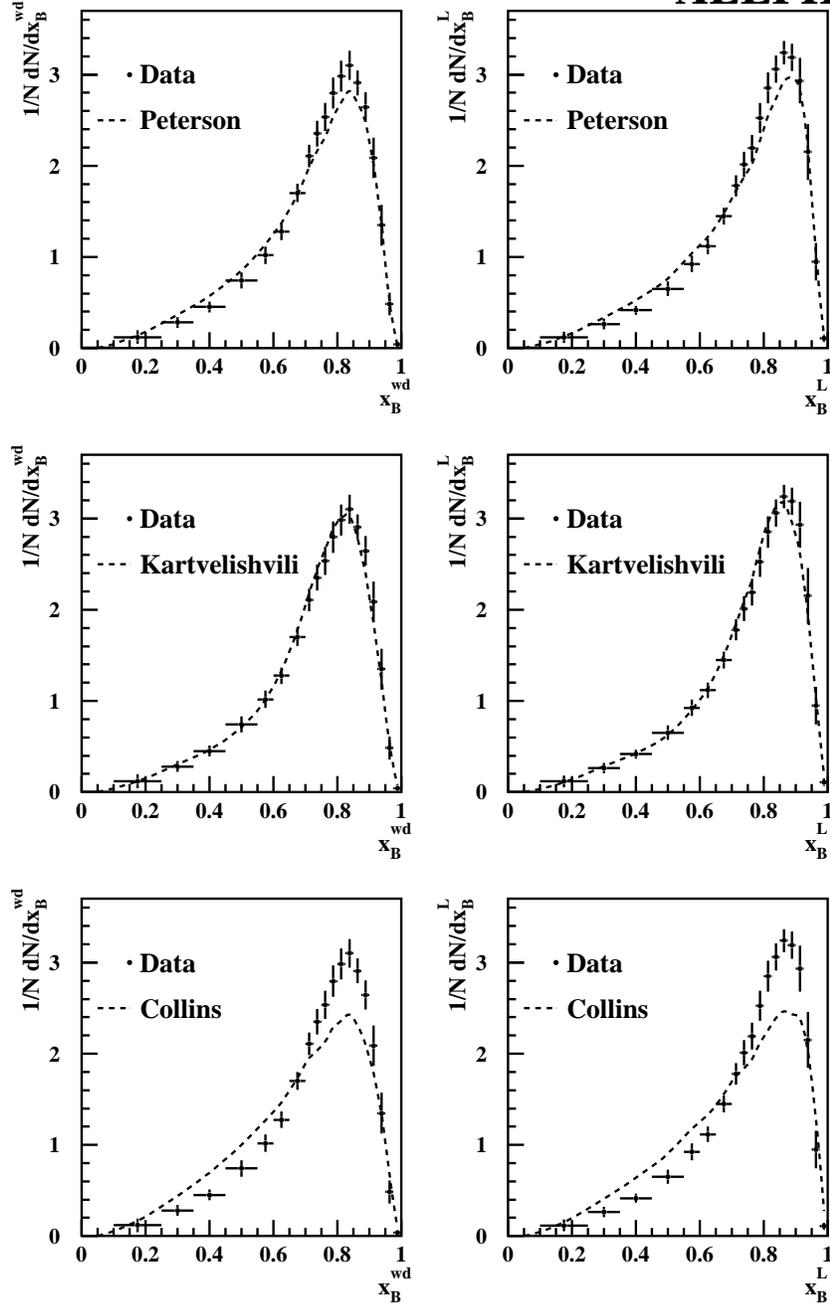}
    \caption{\footnotesize Scaled energy of the leading and weakly-decaying $B$ meson, as reconstructed from data. 
The best-fit distributions for the Peterson model, the Kartvelishvili model, 
and the Collins model are superimposed. For the data, the bin-to-bin errors are highly correlated, as shown in the error matrices in the Appendix.}
    \label{fig:xewd}
  \end{center}
\end{figure}

\section{Systematic checks}
\label{sec:checks}
Possible systematic effects intrinsic to the analysis method are
checked by measuring the energy spectra in a sample of 8 million
simulated $q\bar q$ events. The average values for the scaled
energies of the weakly-decaying and leading $B$ mesons are measured
to be:

\begin{eqnarray}
  \nonumber    \mxel_{\mathrm{MC}}  & = & 0.711 \; \pm 0.005  \, \mathrm{(stat)} \quad ,\\
  \nonumber    \mxewd_{\mathrm{MC}} & = & 0.692 \; \pm 0.005  \, \mathrm{(stat)} \quad ,
\end{eqnarray}
which compare well with the true values:
\begin{eqnarray}
  \nonumber    \mxel_{\mathrm{MC}} ^{\mathrm{true}}&= & 0.712 \quad , \\
  \nonumber   \mxewd_{\mathrm{MC}} ^{\mathrm{true}}&= & 0.692 \quad .
\end{eqnarray}

Electron and muon identification are affected by
different sources of background, and the selection efficiencies and
purities have a different dependence upon the track kinematics and
isolation.  It is therefore interesting to perform the analysis using
separately events with electron candidates or muon candidates.
Consistent results within uncorrelated errors are found:

\begin{equation}
\begin{array}{rr}
   \mxewd_{\mathrm{electrons}}  =  0.724 \; \pm 0.010 \quad ,&  \mxel_{\mathrm{electrons}}  =  0.743 \; \pm 0.010 \quad,\\
   \mxewd_{\mathrm{muons}}  =  0.700 \; \pm 0.014 \quad ,&  \mxel_{\mathrm{muons}}  =  0.720 \; \pm 0.014 \quad.
\end{array}
\end{equation}

The acceptance corrections for the five channels are significantly
different (Fig.~\ref{fig:acceptances}) and the same is true for the resolution
matrices. An inaccurate description of these inputs would easily lead
to incompatible results among the different channels. This is checked
by performing the analysis separately in the five sub-samples. The
results, reported in Table~\ref{tab:separate}, are compatible within
 uncorrelated uncertainties.

\begin{table}[htbp]
  \begin{center}
    \begin{tabular}{|c|c|c|}
      \hline \rule{0pt}{4.8mm}
      Channel & $\mxewd$& $\mxel$ \\ [0.15cm]
      \hline 
      1& $0.700 \pm 0.015$&$0.720 \pm 0.016$  \\
      2& $0.700 \pm 0.020$&$0.720 \pm 0.022$  \\ 
      3& $0.714 \pm 0.012$&$0.733 \pm 0.013$  \\
      4& $0.720 \pm 0.019$&$0.740 \pm 0.019$  \\
      5& $0.738 \pm 0.012$&$0.755 \pm 0.013$  \\\hline
    \end{tabular}
    \caption{\footnotesize Results using the five channels separately. The errors are 
uncorrelated.}
    \label{tab:separate}
  \end{center}
\end{table}

It has been checked that the results are independent of the choice of fragmentation functions in the Monte Carlo sample used to estimate the resolution matrix $G_{ij}$ and the acceptances $\epsilon_{i}$.

As explained in Section~\ref{sec:boh}, the weights applied to reweight
$G_{ij}$ to a given fragmentation function are smoothed with a
polynomial function to reduce the bin-to-bin fluctuations.  However,
the values for the mean scaled energies move by a small fraction of
the statistical errors when such smoothing is not applied, and the
total statistical errors remain nearly constant:
\begin{equation}
\nonumber \mxewd = 0.7177\pm 0.0060\quad , \quad \mxel = 0.7370\pm 0.0065 \quad.
\end{equation}

Heavy flavoured hadrons originating from gluon splitting $g\to b \bar
b$ have an energy much lower than hadrons coming from primary $b$
quarks.  A check on Monte Carlo events shows that the contribution of
such events is negligible.
 
The analysis uses a binned representation of the fragmentation
functions to compensate the relatively small statistical sample in the
data. The binning chosen must not introduce biases in the measured
values nor should it affect the statistical errors. This is checked by
performing a number of analyses in which the binning is varied
randomly around the standard one. Both the central values and the
statistical uncertainties are stable.

\section{Conclusions}

Using the data collected by the ALEPH experiment at and around the $Z$
resonance in the years 1991--1995, about 3400 semileptonic $B^0$ and
$B^\pm$ decays have been selected.  The scaled energy spectra of
weakly-decaying and leading $B$ mesons have been reconstructed, and
their mean values were found to be:

\begin{eqnarray*}
\mxewd &=& 0.7163 \; \pm 0.0061  \, (\mathrm{stat}) \; \pm 0.0056 \, (\mathrm{syst}) \quad , \\
\mxel  &=& 0.7361 \; \pm 0.0063  \, (\mathrm{stat}) \; \pm 0.0063 \, (\mathrm{syst}) \quad . 
\end{eqnarray*}

The observed spectra have been compared with the prediction of JETSET
7.4 using different fragmentation models.  The models of Peterson
\etal~\cite{peterson} and Kartvelishvili \etal~\cite{kart} give a
reasonable description of the data while the Collins
\etal~\cite{collins} model is clearly disfavoured.

This measurement supersedes a previous analysis from
ALEPH~\cite{oldaleph}, which used a different method and smaller
statistics.

The present result  is compatible  with the published results using $b$ hadrons
from L3~\cite{l3ridicolo}, OPAL~\cite{OPAL2} and SLD~\cite{SLD}, and using $B$ mesons from
OPAL~\cite{OPAL1}.

\section{Acknowledgements}
We wish to thank our colleagues in the CERN accelerator divisions for
the successful operation of LEP. We are indebted to the engineers and
technicians in all our institutions for their contribution to the
excellent performance of ALEPH. Those of us from non-member countries
thank CERN for its hospitality.

\appendix

\section{Appendix}

Tables~\ref{tab:wdstat} and~\ref{tab:wdtotal} show the statistical and total error matrices for $\xewd$; 
Tables~\ref{tab:lstat} and~\ref{tab:ltotal} give the same information for $\xel$.

 \begin{table}[!htbp]
 \begin{center}
 \begin{sideways}
 \begin{minipage}[b]{\textheight}
 \begin{center}
 \footnotesize
 \begin{tabular}{|r|r|r|r|r|r|r|r|r|r|r|r|r|r|r|r|r|r|r|r|}
 \hline
\multicolumn{1}{|c|}{bin} &
\multicolumn{1}{c|}{2} &
\multicolumn{1}{c|}{3} &
\multicolumn{1}{c|}{4} &
\multicolumn{1}{c|}{5} &
\multicolumn{1}{c|}{6} &
\multicolumn{1}{c|}{7} &
\multicolumn{1}{c|}{8} &
\multicolumn{1}{c|}{9} &
\multicolumn{1}{c|}{10} &
\multicolumn{1}{c|}{11} &
\multicolumn{1}{c|}{12} &
\multicolumn{1}{c|}{13} &
\multicolumn{1}{c|}{14} &
\multicolumn{1}{c|}{15} &
\multicolumn{1}{c|}{16} &
\multicolumn{1}{c|}{17} &
\multicolumn{1}{c|}{18} &
\multicolumn{1}{c|}{19} &
\multicolumn{1}{c|}{20} \\
\hline
$ 2$ &$ 53.$&$ 28.$&$ -3.$&$-31.$&$-18.$&$-15.$&$ -9.$&$ -1.$&$  2.$&$  5.$&$  7.$&$  7.$&$  6.$&$  2.$&$ -3.$&$ -9.$&$-12.$&$ -8.$&$ -1.  $  \\
$ 3$ &$ 28.$&$ 22.$&$  8.$&$-10.$&$-10.$&$-11.$&$-10.$&$ -4.$&$ -3.$&$ -1.$&$  0.$&$  1.$&$  1.$&$  1.$&$ -1.$&$ -3.$&$ -4.$&$ -3.$&$  0.  $  \\
$ 4$ &$ -3.$&$  8.$&$ 15.$&$ 14.$&$  3.$&$ -1.$&$ -6.$&$ -5.$&$ -6.$&$ -7.$&$ -7.$&$ -7.$&$ -5.$&$ -3.$&$  0.$&$  3.$&$  4.$&$  3.$&$  1.  $  \\
$ 5$ &$-31.$&$-10.$&$ 14.$&$ 35.$&$ 17.$&$ 12.$&$  3.$&$ -3.$&$ -6.$&$ -9.$&$-12.$&$-13.$&$-12.$&$ -8.$&$ -2.$&$  6.$&$ 10.$&$  7.$&$  1.  $  \\
$ 6$ &$-18.$&$-10.$&$  3.$&$ 17.$&$ 11.$&$ 10.$&$  6.$&$  1.$&$ -1.$&$ -3.$&$ -5.$&$ -6.$&$ -6.$&$ -5.$&$ -2.$&$  1.$&$  4.$&$  3.$&$  1.  $  \\
$ 7$ &$-15.$&$-11.$&$ -1.$&$ 12.$&$ 10.$&$ 10.$&$  9.$&$  3.$&$  1.$&$ -1.$&$ -2.$&$ -4.$&$ -5.$&$ -5.$&$ -3.$&$ -1.$&$  1.$&$  2.$&$  0.  $  \\
$ 8$ &$ -9.$&$-10.$&$ -6.$&$  3.$&$  6.$&$  9.$&$ 11.$&$  5.$&$  4.$&$  3.$&$  2.$&$  0.$&$ -2.$&$ -3.$&$ -4.$&$ -4.$&$ -3.$&$ -1.$&$  0.  $  \\
$ 9$ &$ -1.$&$ -4.$&$ -5.$&$ -3.$&$  1.$&$  3.$&$  5.$&$  4.$&$  4.$&$  3.$&$  3.$&$  2.$&$  1.$&$ -1.$&$ -2.$&$ -3.$&$ -3.$&$ -2.$&$  0.  $  \\
$10$ &$  2.$&$ -3.$&$ -6.$&$ -6.$&$ -1.$&$  1.$&$  4.$&$  4.$&$  5.$&$  5.$&$  5.$&$  4.$&$  2.$&$  0.$&$ -2.$&$ -4.$&$ -5.$&$ -3.$&$  0.  $  \\
$11$ &$  5.$&$ -1.$&$ -7.$&$ -9.$&$ -3.$&$ -1.$&$  3.$&$  3.$&$  5.$&$  6.$&$  6.$&$  5.$&$  4.$&$  1.$&$ -2.$&$ -4.$&$ -5.$&$ -3.$&$ -1.  $  \\
$12$ &$  7.$&$  0.$&$ -7.$&$-12.$&$ -5.$&$ -2.$&$  2.$&$  3.$&$  5.$&$  6.$&$  7.$&$  7.$&$  5.$&$  2.$&$ -1.$&$ -4.$&$ -6.$&$ -4.$&$ -1.  $  \\
$13$ &$  7.$&$  1.$&$ -7.$&$-13.$&$ -6.$&$ -4.$&$  0.$&$  2.$&$  4.$&$  5.$&$  7.$&$  7.$&$  6.$&$  4.$&$  0.$&$ -3.$&$ -5.$&$ -4.$&$ -1.  $  \\
$14$ &$  6.$&$  1.$&$ -5.$&$-12.$&$ -6.$&$ -5.$&$ -2.$&$  1.$&$  2.$&$  4.$&$  5.$&$  6.$&$  6.$&$  5.$&$  2.$&$ -1.$&$ -3.$&$ -3.$&$  0.  $  \\
$15$ &$  2.$&$  1.$&$ -3.$&$ -8.$&$ -5.$&$ -5.$&$ -3.$&$ -1.$&$  0.$&$  1.$&$  2.$&$  4.$&$  5.$&$  5.$&$  4.$&$  2.$&$  0.$&$ -1.$&$  0.  $  \\
$16$ &$ -3.$&$ -1.$&$  0.$&$ -2.$&$ -2.$&$ -3.$&$ -4.$&$ -2.$&$ -2.$&$ -2.$&$ -1.$&$  0.$&$  2.$&$  4.$&$  5.$&$  5.$&$  4.$&$  2.$&$  0.  $  \\
$17$ &$ -9.$&$ -3.$&$  3.$&$  6.$&$  1.$&$ -1.$&$ -4.$&$ -3.$&$ -4.$&$ -4.$&$ -4.$&$ -3.$&$ -1.$&$  2.$&$  5.$&$  8.$&$  8.$&$  5.$&$  1.  $  \\
$18$ &$-12.$&$ -4.$&$  4.$&$ 10.$&$  4.$&$  1.$&$ -3.$&$ -3.$&$ -5.$&$ -5.$&$ -6.$&$ -5.$&$ -3.$&$  0.$&$  4.$&$  8.$&$ 10.$&$  6.$&$  1.  $  \\
$19$ &$ -8.$&$ -3.$&$  3.$&$  7.$&$  3.$&$  2.$&$ -1.$&$ -2.$&$ -3.$&$ -3.$&$ -4.$&$ -4.$&$ -3.$&$ -1.$&$  2.$&$  5.$&$  6.$&$  4.$&$  1.  $  \\
$20$ &$ -1.$&$  0.$&$  1.$&$  1.$&$  1.$&$  0.$&$  0.$&$  0.$&$  0.$&$ -1.$&$ -1.$&$ -1.$&$  0.$&$  0.$&$  0.$&$  1.$&$  1.$&$  1.$&$  0.  $  \\
 \hline
 \end{tabular}
 \caption{\footnotesize Statistical error matrix for $\xewd$. All the numbers are in units of $10^{-6}$.}

 \label{tab:wdstat}
 \end{center}
 \end{minipage}
 \end{sideways}
 \end{center}
 \end{table}
 \begin{table}[!htbp]
 \begin{center}
 \begin{sideways}
 \begin{minipage}[b]{\textheight}
 \begin{center}
 \footnotesize
 \begin{tabular}{|r|r|r|r|r|r|r|r|r|r|r|r|r|r|r|r|r|r|r|r|}
 \hline
\multicolumn{1}{|c|}{bin} &
\multicolumn{1}{c|}{2} &
\multicolumn{1}{c|}{3} &
\multicolumn{1}{c|}{4} &
\multicolumn{1}{c|}{5} &
\multicolumn{1}{c|}{6} &
\multicolumn{1}{c|}{7} &
\multicolumn{1}{c|}{8} &
\multicolumn{1}{c|}{9} &
\multicolumn{1}{c|}{10} &
\multicolumn{1}{c|}{11} &
\multicolumn{1}{c|}{12} &
\multicolumn{1}{c|}{13} &
\multicolumn{1}{c|}{14} &
\multicolumn{1}{c|}{15} &
\multicolumn{1}{c|}{16} &
\multicolumn{1}{c|}{17} &
\multicolumn{1}{c|}{18} &
\multicolumn{1}{c|}{19} &
\multicolumn{1}{c|}{20} \\
\hline
$ 2$ &$127.$&$ 42.$&$-28.$&$-72.$&$-32.$&$-18.$&$  1.$&$ 10.$&$ 14.$&$ 18.$&$ 19.$&$ 15.$&$  7.$&$ -3.$&$-17.$&$-30.$&$-31.$&$-18.$&$ -3.  $  \\
$ 3$ &$ 42.$&$ 34.$&$ 10.$&$-23.$&$-18.$&$-18.$&$-13.$&$ -3.$&$  0.$&$  3.$&$  6.$&$  7.$&$  6.$&$  3.$&$ -3.$&$-10.$&$-13.$&$ -8.$&$ -2.  $  \\
$ 4$ &$-28.$&$ 10.$&$ 34.$&$ 33.$&$  7.$&$ -2.$&$-14.$&$-11.$&$-13.$&$-13.$&$-13.$&$-10.$&$ -7.$&$ -1.$&$  4.$&$  9.$&$  9.$&$  6.$&$  1.  $  \\
$ 5$ &$-72.$&$-23.$&$ 33.$&$ 79.$&$ 37.$&$ 24.$&$  1.$&$-11.$&$-19.$&$-25.$&$-30.$&$-30.$&$-25.$&$-14.$&$  3.$&$ 21.$&$ 29.$&$ 18.$&$  3.  $  \\
$ 6$ &$-32.$&$-18.$&$  7.$&$ 37.$&$ 22.$&$ 19.$&$ 10.$&$ -1.$&$ -5.$&$-10.$&$-13.$&$-15.$&$-15.$&$-10.$&$ -3.$&$  6.$&$ 12.$&$  8.$&$  2.  $  \\
$ 7$ &$-18.$&$-18.$&$ -2.$&$ 24.$&$ 19.$&$ 21.$&$ 17.$&$  5.$&$  1.$&$ -3.$&$ -7.$&$-10.$&$-13.$&$-11.$&$ -8.$&$ -2.$&$  3.$&$  3.$&$  1.  $  \\
$ 8$ &$  1.$&$-13.$&$-14.$&$  1.$&$ 10.$&$ 17.$&$ 25.$&$ 13.$&$ 11.$&$  9.$&$  5.$&$  0.$&$ -6.$&$-11.$&$-15.$&$-16.$&$-12.$&$ -5.$&$ -1.  $  \\
$ 9$ &$ 10.$&$ -3.$&$-11.$&$-11.$&$ -1.$&$  5.$&$ 13.$&$  9.$&$ 10.$&$ 10.$&$  9.$&$  6.$&$  1.$&$ -3.$&$ -9.$&$-13.$&$-13.$&$ -7.$&$ -1.  $  \\
$10$ &$ 14.$&$  0.$&$-13.$&$-19.$&$ -5.$&$  1.$&$ 11.$&$ 10.$&$ 12.$&$ 13.$&$ 13.$&$ 10.$&$  5.$&$ -1.$&$ -9.$&$-16.$&$-16.$&$ -9.$&$ -2.  $  \\
$11$ &$ 18.$&$  3.$&$-13.$&$-25.$&$-10.$&$ -3.$&$  9.$&$ 10.$&$ 13.$&$ 15.$&$ 16.$&$ 14.$&$  9.$&$  2.$&$ -8.$&$-17.$&$-19.$&$-11.$&$ -2.  $  \\
$12$ &$ 19.$&$  6.$&$-13.$&$-30.$&$-13.$&$ -7.$&$  5.$&$  9.$&$ 13.$&$ 16.$&$ 18.$&$ 17.$&$ 13.$&$  5.$&$ -6.$&$-17.$&$-20.$&$-12.$&$ -2.  $  \\
$13$ &$ 15.$&$  7.$&$-10.$&$-30.$&$-15.$&$-10.$&$  0.$&$  6.$&$ 10.$&$ 14.$&$ 17.$&$ 18.$&$ 15.$&$  8.$&$ -2.$&$-13.$&$-17.$&$-10.$&$ -2.  $  \\
$14$ &$  7.$&$  6.$&$ -7.$&$-25.$&$-15.$&$-13.$&$ -6.$&$  1.$&$  5.$&$  9.$&$ 13.$&$ 15.$&$ 16.$&$ 11.$&$  4.$&$ -5.$&$-10.$&$ -7.$&$ -1.  $  \\
$15$ &$ -3.$&$  3.$&$ -1.$&$-14.$&$-10.$&$-11.$&$-11.$&$ -3.$&$ -1.$&$  2.$&$  5.$&$  8.$&$ 11.$&$ 11.$&$ 10.$&$  6.$&$  1.$&$ -1.$&$  0.  $  \\
$16$ &$-17.$&$ -3.$&$  4.$&$  3.$&$ -3.$&$ -8.$&$-15.$&$ -9.$&$ -9.$&$ -8.$&$ -6.$&$ -2.$&$  4.$&$ 10.$&$ 17.$&$ 19.$&$ 16.$&$  8.$&$  1.  $  \\
$17$ &$-30.$&$-10.$&$  9.$&$ 21.$&$  6.$&$ -2.$&$-16.$&$-13.$&$-16.$&$-17.$&$-17.$&$-13.$&$ -5.$&$  6.$&$ 19.$&$ 30.$&$ 30.$&$ 16.$&$  3.  $  \\
$18$ &$-31.$&$-13.$&$  9.$&$ 29.$&$ 12.$&$  3.$&$-12.$&$-13.$&$-16.$&$-19.$&$-20.$&$-17.$&$-10.$&$  1.$&$ 16.$&$ 30.$&$ 32.$&$ 17.$&$  3.  $  \\
$19$ &$-18.$&$ -8.$&$  6.$&$ 18.$&$  8.$&$  3.$&$ -5.$&$ -7.$&$ -9.$&$-11.$&$-12.$&$-10.$&$ -7.$&$ -1.$&$  8.$&$ 16.$&$ 17.$&$ 10.$&$  2.  $  \\
$20$ &$ -3.$&$ -2.$&$  1.$&$  3.$&$  2.$&$  1.$&$ -1.$&$ -1.$&$ -2.$&$ -2.$&$ -2.$&$ -2.$&$ -1.$&$  0.$&$  1.$&$  3.$&$  3.$&$  2.$&$  0.  $  \\
 \hline
 \end{tabular}
 \caption{\footnotesize Total error matrix for $\xewd$. All the numbers are in units of $10^{-6}$.}
 \label{tab:wdtotal}
 \end{center}
 \end{minipage}
 \end{sideways}
 \end{center}
 \end{table}

 \begin{table}[!htbp]
 \begin{center}
 \begin{sideways}
 \begin{minipage}[b]{\textheight}
 \begin{center}
 \footnotesize
 \begin{tabular}{|r|r|r|r|r|r|r|r|r|r|r|r|r|r|r|r|r|r|r|r|}
 \hline
\multicolumn{1}{|c|}{bin} &
\multicolumn{1}{c|}{2} &
\multicolumn{1}{c|}{3} &
\multicolumn{1}{c|}{4} &
\multicolumn{1}{c|}{5} &
\multicolumn{1}{c|}{6} &
\multicolumn{1}{c|}{7} &
\multicolumn{1}{c|}{8} &
\multicolumn{1}{c|}{9} &
\multicolumn{1}{c|}{10} &
\multicolumn{1}{c|}{11} &
\multicolumn{1}{c|}{12} &
\multicolumn{1}{c|}{13} &
\multicolumn{1}{c|}{14} &
\multicolumn{1}{c|}{15} &
\multicolumn{1}{c|}{16} &
\multicolumn{1}{c|}{17} &
\multicolumn{1}{c|}{18} &
\multicolumn{1}{c|}{19} &
\multicolumn{1}{c|}{20} \\
\hline
$2$ &$ 49.$&$ 28.$&$  0.$&$-26.$&$-17.$&$-14.$&$ -9.$&$ -2.$&$  1.$&$  3.$&$  6.$&$  7.$&$  7.$&$  5.$&$ -1.$&$ -8.$&$-14.$&$-11.$&$ -2.  $  \\
$3$ &$ 28.$&$ 21.$&$  8.$&$ -8.$&$ -9.$&$ -9.$&$ -9.$&$ -4.$&$ -3.$&$ -2.$&$ -1.$&$  0.$&$  1.$&$  0.$&$ -1.$&$ -3.$&$ -5.$&$ -4.$&$ -1.  $  \\
$4$ &$  0.$&$  8.$&$ 14.$&$ 13.$&$  3.$&$  0.$&$ -5.$&$ -4.$&$ -5.$&$ -6.$&$ -7.$&$ -7.$&$ -7.$&$ -5.$&$ -2.$&$  1.$&$  4.$&$  4.$&$  1.  $  \\
$5$ &$-26.$&$ -8.$&$ 13.$&$ 30.$&$ 15.$&$ 11.$&$  4.$&$ -2.$&$ -5.$&$ -7.$&$-10.$&$-12.$&$-12.$&$-10.$&$ -4.$&$  3.$&$ 10.$&$  9.$&$  2.  $  \\
$6$ &$-17.$&$ -9.$&$  3.$&$ 15.$&$ 10.$&$  8.$&$  5.$&$  1.$&$ -1.$&$ -2.$&$ -4.$&$ -5.$&$ -6.$&$ -5.$&$ -3.$&$  0.$&$  4.$&$  4.$&$  1.  $  \\
$7$ &$-14.$&$ -9.$&$  0.$&$ 11.$&$  8.$&$  8.$&$  7.$&$  2.$&$  1.$&$  0.$&$ -2.$&$ -3.$&$ -4.$&$ -4.$&$ -3.$&$ -1.$&$  1.$&$  2.$&$  0.  $  \\
$8$ &$ -9.$&$ -9.$&$ -5.$&$  4.$&$  5.$&$  7.$&$  9.$&$  4.$&$  4.$&$  3.$&$  2.$&$  1.$&$ -1.$&$ -2.$&$ -3.$&$ -4.$&$ -3.$&$ -2.$&$  0.  $  \\
$9$ &$ -2.$&$ -4.$&$ -4.$&$ -2.$&$  1.$&$  2.$&$  4.$&$  3.$&$  3.$&$  3.$&$  3.$&$  2.$&$  1.$&$  0.$&$ -2.$&$ -3.$&$ -4.$&$ -3.$&$ -1.  $  \\
$10$&$  1.$&$ -3.$&$ -5.$&$ -5.$&$ -1.$&$  1.$&$  4.$&$  3.$&$  4.$&$  4.$&$  4.$&$  4.$&$  3.$&$  1.$&$ -1.$&$ -4.$&$ -5.$&$ -4.$&$ -1.  $  \\
$11$&$  3.$&$ -2.$&$ -6.$&$ -7.$&$ -2.$&$  0.$&$  3.$&$  3.$&$  4.$&$  5.$&$  5.$&$  5.$&$  4.$&$  2.$&$ -1.$&$ -4.$&$ -6.$&$ -5.$&$ -1.  $  \\
$12$&$  6.$&$ -1.$&$ -7.$&$-10.$&$ -4.$&$ -2.$&$  2.$&$  3.$&$  4.$&$  5.$&$  6.$&$  7.$&$  6.$&$  4.$&$  0.$&$ -4.$&$ -7.$&$ -6.$&$ -1.  $  \\
$13$&$  7.$&$  0.$&$ -7.$&$-12.$&$ -5.$&$ -3.$&$  1.$&$  2.$&$  4.$&$  5.$&$  7.$&$  7.$&$  7.$&$  5.$&$  1.$&$ -4.$&$ -7.$&$ -6.$&$ -1.  $  \\
$14$&$  7.$&$  1.$&$ -7.$&$-12.$&$ -6.$&$ -4.$&$ -1.$&$  1.$&$  3.$&$  4.$&$  6.$&$  7.$&$  7.$&$  5.$&$  2.$&$ -2.$&$ -6.$&$ -5.$&$ -1.  $  \\
$15$&$  5.$&$  0.$&$ -5.$&$-10.$&$ -5.$&$ -4.$&$ -2.$&$  0.$&$  1.$&$  2.$&$  4.$&$  5.$&$  5.$&$  5.$&$  3.$&$  1.$&$ -2.$&$ -2.$&$ -1.  $  \\
$16$&$ -1.$&$ -1.$&$ -2.$&$ -4.$&$ -3.$&$ -3.$&$ -3.$&$ -2.$&$ -1.$&$ -1.$&$  0.$&$  1.$&$  2.$&$  3.$&$  4.$&$  5.$&$  4.$&$  2.$&$  0.  $  \\
$17$&$ -8.$&$ -3.$&$  1.$&$  3.$&$  0.$&$ -1.$&$ -4.$&$ -3.$&$ -4.$&$ -4.$&$ -4.$&$ -4.$&$ -2.$&$  1.$&$  5.$&$  9.$&$ 11.$&$  7.$&$  1.  $  \\
$18$&$-14.$&$ -5.$&$  4.$&$ 10.$&$  4.$&$  1.$&$ -3.$&$ -4.$&$ -5.$&$ -6.$&$ -7.$&$ -7.$&$ -6.$&$ -2.$&$  4.$&$ 11.$&$ 15.$&$ 10.$&$  2.  $  \\
$19$&$-11.$&$ -4.$&$  4.$&$  9.$&$  4.$&$  2.$&$ -2.$&$ -3.$&$ -4.$&$ -5.$&$ -6.$&$ -6.$&$ -5.$&$ -2.$&$  2.$&$  7.$&$ 10.$&$  8.$&$  2.  $  \\
$20$&$ -2.$&$ -1.$&$  1.$&$  2.$&$  1.$&$  0.$&$  0.$&$ -1.$&$ -1.$&$ -1.$&$ -1.$&$ -1.$&$ -1.$&$ -1.$&$  0.$&$  1.$&$  2.$&$  2.$&$  0.  $  \\
 \hline
 \end{tabular}
 \caption{\footnotesize Statistical error matrix for $\xel$. All the numbers are in units of $10^{-6}$.}
 \label{tab:lstat}
 \end{center}
 \end{minipage}
 \end{sideways}
 \end{center}
 \end{table}

 \begin{table}[!htbp]
 \begin{center}
 \begin{sideways}
 \begin{minipage}[b]{\textheight}
 \begin{center}
 \footnotesize
 \begin{tabular}{|r|r|r|r|r|r|r|r|r|r|r|r|r|r|r|r|r|r|r|r|}
 \hline
\multicolumn{1}{|c|}{bin} &
\multicolumn{1}{c|}{2} &
\multicolumn{1}{c|}{3} &
\multicolumn{1}{c|}{4} &
\multicolumn{1}{c|}{5} &
\multicolumn{1}{c|}{6} &
\multicolumn{1}{c|}{7} &
\multicolumn{1}{c|}{8} &
\multicolumn{1}{c|}{9} &
\multicolumn{1}{c|}{10} &
\multicolumn{1}{c|}{11} &
\multicolumn{1}{c|}{12} &
\multicolumn{1}{c|}{13} &
\multicolumn{1}{c|}{14} &
\multicolumn{1}{c|}{15} &
\multicolumn{1}{c|}{16} &
\multicolumn{1}{c|}{17} &
\multicolumn{1}{c|}{18} &
\multicolumn{1}{c|}{19} &
\multicolumn{1}{c|}{20} \\
\hline
$ 2$&$ 90.$&$ 48.$&$ -6.$&$-55.$&$-33.$&$-26.$&$-12.$&$  2.$&$  9.$&$ 14.$&$ 19.$&$ 22.$&$ 20.$&$ 11.$&$ -2.$&$-26.$&$-40.$&$-29.$&$ -6.  $  \\
$ 3$&$ 48.$&$ 34.$&$  9.$&$-19.$&$-17.$&$-16.$&$-12.$&$ -4.$&$ -2.$&$  1.$&$  3.$&$  5.$&$  5.$&$  2.$&$ -1.$&$-10.$&$-14.$&$-11.$&$ -2.  $  \\
$ 4$&$ -6.$&$  9.$&$ 23.$&$ 25.$&$  7.$&$  2.$&$ -7.$&$ -8.$&$-10.$&$-12.$&$-13.$&$-14.$&$-13.$&$ -9.$&$ -3.$&$  6.$&$ 12.$&$  9.$&$  2.  $  \\
$5 $&$-55.$&$-19.$&$ 25.$&$ 62.$&$ 32.$&$ 23.$&$  7.$&$ -5.$&$-12.$&$-17.$&$-23.$&$-27.$&$-26.$&$-20.$&$ -9.$&$ 12.$&$ 27.$&$ 21.$&$  5.  $  \\
$ 6$&$-33.$&$-17.$&$  7.$&$ 32.$&$ 20.$&$ 17.$&$ 11.$&$  2.$&$ -2.$&$ -5.$&$ -9.$&$-12.$&$-13.$&$-11.$&$ -8.$&$  1.$&$  8.$&$  8.$&$  2.  $  \\
$ 7$&$-26.$&$-16.$&$  2.$&$ 23.$&$ 17.$&$ 17.$&$ 15.$&$  6.$&$  3.$&$  1.$&$ -2.$&$ -5.$&$ -8.$&$-10.$&$-10.$&$ -6.$&$ -2.$&$  1.$&$  1.  $  \\
$ 8$&$-12.$&$-12.$&$ -7.$&$  7.$&$ 11.$&$ 15.$&$ 22.$&$ 12.$&$ 12.$&$ 11.$&$  9.$&$  6.$&$  0.$&$ -6.$&$-13.$&$-20.$&$-21.$&$-12.$&$ -2.  $  \\
$ 9$&$  2.$&$ -4.$&$ -8.$&$ -5.$&$  2.$&$  6.$&$ 12.$&$  9.$&$ 10.$&$ 10.$&$ 10.$&$  8.$&$  4.$&$ -1.$&$ -8.$&$-15.$&$-18.$&$-11.$&$ -2.  $  \\
$10$&$  9.$&$ -2.$&$-10.$&$-12.$&$ -2.$&$  3.$&$ 12.$&$ 10.$&$ 12.$&$ 12.$&$ 13.$&$ 12.$&$  8.$&$  1.$&$ -7.$&$-18.$&$-23.$&$-15.$&$ -3.  $  \\
$11$&$ 14.$&$  1.$&$-12.$&$-17.$&$ -5.$&$  1.$&$ 11.$&$ 10.$&$ 12.$&$ 14.$&$ 15.$&$ 14.$&$ 10.$&$  3.$&$ -6.$&$-19.$&$-25.$&$-17.$&$ -3.  $  \\
$12$&$ 19.$&$  3.$&$-13.$&$-23.$&$ -9.$&$ -2.$&$  9.$&$ 10.$&$ 13.$&$ 15.$&$ 17.$&$ 17.$&$ 13.$&$  6.$&$ -5.$&$-20.$&$-28.$&$-19.$&$ -4.  $  \\
$13$&$ 22.$&$  5.$&$-14.$&$-27.$&$-12.$&$ -5.$&$  6.$&$  8.$&$ 12.$&$ 14.$&$ 17.$&$ 19.$&$ 15.$&$  8.$&$ -2.$&$-18.$&$-26.$&$-19.$&$ -4.  $  \\
$14$&$ 20.$&$  5.$&$-13.$&$-26.$&$-13.$&$ -8.$&$  0.$&$  4.$&$  8.$&$ 10.$&$ 13.$&$ 15.$&$ 14.$&$ 10.$&$  3.$&$ -9.$&$-17.$&$-13.$&$ -3.  $  \\
$15$&$ 11.$&$  2.$&$ -9.$&$-20.$&$-11.$&$-10.$&$ -6.$&$ -1.$&$  1.$&$  3.$&$  6.$&$  8.$&$ 10.$&$ 10.$&$  8.$&$  3.$&$ -2.$&$ -3.$&$ -1.  $  \\
$16$&$ -2.$&$ -1.$&$ -3.$&$ -9.$&$ -8.$&$-10.$&$-13.$&$ -8.$&$ -7.$&$ -6.$&$ -5.$&$ -2.$&$  3.$&$  8.$&$ 14.$&$ 19.$&$ 19.$&$ 10.$&$  2.  $  \\
$17$&$-26.$&$-10.$&$  6.$&$ 12.$&$  1.$&$ -6.$&$-20.$&$-15.$&$-18.$&$-19.$&$-20.$&$-18.$&$ -9.$&$  3.$&$ 19.$&$ 39.$&$ 46.$&$ 30.$&$  5.  $  \\
$18$&$-40.$&$-14.$&$ 12.$&$ 27.$&$  8.$&$ -2.$&$-21.$&$-18.$&$-23.$&$-25.$&$-28.$&$-26.$&$-17.$&$ -2.$&$ 19.$&$ 46.$&$ 59.$&$ 39.$&$  7.  $  \\
$19$&$-29.$&$-11.$&$  9.$&$ 21.$&$  8.$&$  1.$&$-12.$&$-11.$&$-15.$&$-17.$&$-19.$&$-19.$&$-13.$&$ -3.$&$ 10.$&$ 30.$&$ 39.$&$ 27.$&$  5.  $  \\
$20$&$ -6.$&$ -2.$&$  2.$&$  5.$&$  2.$&$  1.$&$ -2.$&$ -2.$&$ -3.$&$ -3.$&$ -4.$&$ -4.$&$ -3.$&$ -1.$&$  2.$&$  5.$&$  7.$&$  5.$&$  1.  $  \\
 \hline
 \end{tabular}
 \caption{\footnotesize Total error matrix for $\xel$. All the numbers are in units of $10^{-6}$.}

 \label{tab:ltotal}
 \end{center}
 \end{minipage}
 \end{sideways}
 \end{center}
 \end{table}

\end{document}

%% file: authors.tex
\pagestyle{empty}
\newpage
\small
%
\newlength{\saveparskip}
\newlength{\savetextheight}
\newlength{\savetopmargin}
\newlength{\savetextwidth}
\newlength{\saveoddsidemargin}
\newlength{\savetopsep}
\setlength{\saveparskip}{\parskip}
\setlength{\savetextheight}{\textheight}
\setlength{\savetopmargin}{\topmargin}
\setlength{\savetextwidth}{\textwidth}
\setlength{\saveoddsidemargin}{\oddsidemargin}
\setlength{\savetopsep}{\topsep}
%
%
\setlength{\parskip}{0.0cm}
\setlength{\textheight}{25.0cm}
\setlength{\topmargin}{-1.5cm}
\setlength{\textwidth}{16 cm}
\setlength{\oddsidemargin}{-0.0cm}
\setlength{\topsep}{1mm}
\pretolerance=10000
\centerline{\large\bf The ALEPH Collaboration}
\footnotesize
\vspace{0.5cm}
{\raggedbottom
\begin{sloppypar}
\samepage\noindent
A.~Heister,
S.~Schael
\nopagebreak
\begin{center}
\parbox{15.5cm}{\sl\samepage
Physikalisches Institut das RWTH-Aachen, D-52056 Aachen, Germany}
\end{center}\end{sloppypar}
\vspace{2mm}
\begin{sloppypar}
\noindent
R.~Barate,
I.~De~Bonis,
D.~Decamp,
C.~Goy,
\mbox{J.-P.~Lees},
E.~Merle,
\mbox{M.-N.~Minard},
B.~Pietrzyk
\nopagebreak
\begin{center}
\parbox{15.5cm}{\sl\samepage
Laboratoire de Physique des Particules (LAPP), IN$^{2}$P$^{3}$-CNRS,
F-74019 Annecy-le-Vieux Cedex, France}
\end{center}\end{sloppypar}
\vspace{2mm}
\begin{sloppypar}
\noindent
S.~Bravo,
M.P.~Casado,
M.~Chmeissani,
J.M.~Crespo,
E.~Fernandez,
\mbox{M.~Fernandez-Bosman},
Ll.~Garrido,$^{15}$
E.~Graug\'{e}s,
M.~Martinez,
G.~Merino,
R.~Miquel,$^{27}$
Ll.M.~Mir,$^{27}$
A.~Pacheco,
H.~Ruiz
\nopagebreak
\begin{center}
\parbox{15.5cm}{\sl\samepage
Institut de F\'{i}sica d'Altes Energies, Universitat Aut\`{o}noma
de Barcelona, E-08193 Bellaterra (Barcelona), Spain$^{7}$}
\end{center}\end{sloppypar}
\vspace{2mm}
\begin{sloppypar}
\noindent
A.~Colaleo,
D.~Creanza,
M.~de~Palma,
G.~Iaselli,
G.~Maggi,
M.~Maggi,
S.~Nuzzo,
A.~Ranieri,
G.~Raso,$^{23}$
F.~Ruggieri,
G.~Selvaggi,
L.~Silvestris,
P.~Tempesta,
A.~Tricomi,$^{3}$
G.~Zito
\nopagebreak
\begin{center}
\parbox{15.5cm}{\sl\samepage
Dipartimento di Fisica, INFN Sezione di Bari, I-70126
Bari, Italy}
\end{center}\end{sloppypar}
\vspace{2mm}
\begin{sloppypar}
\noindent
X.~Huang,
J.~Lin,
Q. Ouyang,
T.~Wang,
Y.~Xie,
R.~Xu,
S.~Xue,
J.~Zhang,
L.~Zhang,
W.~Zhao
\nopagebreak
\begin{center}
\parbox{15.5cm}{\sl\samepage
Institute of High Energy Physics, Academia Sinica, Beijing, The People's
Republic of China$^{8}$}
\end{center}\end{sloppypar}
\vspace{2mm}
\begin{sloppypar}
\noindent
D.~Abbaneo,
P.~Azzurri,
G.~Boix,$^{6}$
O.~Buchm\"uller,
M.~Cattaneo,
F.~Cerutti,
B.~Clerbaux,
G.~Dissertori,
H.~Drevermann,
R.W.~Forty,
M.~Frank,
T.C.~Greening,
J.B.~Hansen,
J.~Harvey,
P.~Janot,
B.~Jost,
M.~Kado,
P.~Mato,
A.~Moutoussi,
F.~Ranjard,
L.~Rolandi,
D.~Schlatter,
O.~Schneider,$^{2}$
P.~Spagnolo,
W.~Tejessy,
F.~Teubert,
E.~Tournefier,$^{25}$
J.~Ward
\nopagebreak
\begin{center}
\parbox{15.5cm}{\sl\samepage
European Laboratory for Particle Physics (CERN), CH-1211 Geneva 23,
Switzerland}
\end{center}\end{sloppypar}
\vspace{2mm}
\begin{sloppypar}
\noindent
Z.~Ajaltouni,
F.~Badaud,
A.~Falvard,$^{22}$
P.~Gay,
P.~Henrard,
J.~Jousset,
B.~Michel,
S.~Monteil,
\mbox{J-C.~Montret},
D.~Pallin,
P.~Perret,
F.~Podlyski
\nopagebreak
\begin{center}
\parbox{15.5cm}{\sl\samepage
Laboratoire de Physique Corpusculaire, Universit\'e Blaise Pascal,
IN$^{2}$P$^{3}$-CNRS, Clermont-Ferrand, F-63177 Aubi\`{e}re, France}
\end{center}\end{sloppypar}
\vspace{2mm}
\begin{sloppypar}
\noindent
J.D.~Hansen,
J.R.~Hansen,
P.H.~Hansen,
B.S.~Nilsson,
A.~W\"a\"an\"anen
\begin{center}
\parbox{15.5cm}{\sl\samepage
Niels Bohr Institute, DK-2100 Copenhagen, Denmark$^{9}$}
\end{center}\end{sloppypar}
\vspace{2mm}
\begin{sloppypar}
\noindent
A.~Kyriakis,
C.~Markou,
E.~Simopoulou,
A.~Vayaki,
K.~Zachariadou
\nopagebreak
\begin{center}
\parbox{15.5cm}{\sl\samepage
Nuclear Research Center Demokritos (NRCD), GR-15310 Attiki, Greece}
\end{center}\end{sloppypar}
\vspace{2mm}
\begin{sloppypar}
\noindent
A.~Blondel,$^{12}$
G.~Bonneaud,
\mbox{J.-C.~Brient},
A.~Roug\'{e},
M.~Rumpf,
M.~Swynghedauw,
M.~Verderi,
\linebreak
H.~Videau
\nopagebreak
\begin{center}
\parbox{15.5cm}{\sl\samepage
Laboratoire de Physique Nucl\'eaire et des Hautes Energies, Ecole
Polytechnique, IN$^{2}$P$^{3}$-CNRS, \mbox{F-91128} Palaiseau Cedex, France}
\end{center}\end{sloppypar}
\vspace{2mm}
\begin{sloppypar}
\noindent
V.~Ciulli,
E.~Focardi,
G.~Parrini
\nopagebreak
\begin{center}
\parbox{15.5cm}{\sl\samepage
Dipartimento di Fisica, Universit\`a di Firenze, INFN Sezione di Firenze,
I-50125 Firenze, Italy}
\end{center}\end{sloppypar}
\vspace{2mm}
\begin{sloppypar}
\noindent
A.~Antonelli,
M.~Antonelli,
G.~Bencivenni,
G.~Bologna,$^{4}$
F.~Bossi,
P.~Campana,
G.~Capon,
V.~Chiarella,
P.~Laurelli,
G.~Mannocchi,$^{5}$
F.~Murtas,
G.P.~Murtas,
L.~Passalacqua,
\mbox{M.~Pepe-Altarelli}$^{24}$
\nopagebreak
\begin{center}
\parbox{15.5cm}{\sl\samepage
Laboratori Nazionali dell'INFN (LNF-INFN), I-00044 Frascati, Italy}
\end{center}\end{sloppypar}
\vspace{2mm}
\begin{sloppypar}
\noindent
A.W. Halley,
J.G.~Lynch,
P.~Negus,
V.~O'Shea,
C.~Raine,
A.S.~Thompson
\nopagebreak
\begin{center}
\parbox{15.5cm}{\sl\samepage
Department of Physics and Astronomy, University of Glasgow, Glasgow G12
8QQ,United Kingdom$^{10}$}
\end{center}\end{sloppypar}
\vspace{2mm}
\begin{sloppypar}
\noindent
S.~Wasserbaech
\nopagebreak
\begin{center}
\parbox{15.5cm}{\sl\samepage
Department of Physics, Haverford College, Haverford, PA 19041-1392, U.S.A.}
\end{center}\end{sloppypar}
\vspace{2mm}
\begin{sloppypar}
\noindent
R.~Cavanaugh,
S.~Dhamotharan,
C.~Geweniger,
P.~Hanke,
G.~Hansper,
V.~Hepp,
E.E.~Kluge,
A.~Putzer,
J.~Sommer,
K.~Tittel,
S.~Werner,$^{19}$
M.~Wunsch$^{19}$
\nopagebreak
\begin{center}
\parbox{15.5cm}{\sl\samepage
Kirchhoff-Institut f\"ur Physik, Universit\"at Heidelberg, D-69120
Heidelberg, Germany$^{16}$}
\end{center}\end{sloppypar}
\vspace{2mm}
\begin{sloppypar}
\noindent
R.~Beuselinck,
D.M.~Binnie,
W.~Cameron,
P.J.~Dornan,
M.~Girone,$^{1}$
N.~Marinelli,
J.K.~Sedgbeer,
J.C.~Thompson$^{14}$
\nopagebreak
\begin{center}
\parbox{15.5cm}{\sl\samepage
Department of Physics, Imperial College, London SW7 2BZ,
United Kingdom$^{10}$}
\end{center}\end{sloppypar}
\vspace{2mm}
\begin{sloppypar}
\noindent
V.M.~Ghete,
P.~Girtler,
E.~Kneringer,
D.~Kuhn,
G.~Rudolph
\nopagebreak
\begin{center}
\parbox{15.5cm}{\sl\samepage
Institut f\"ur Experimentalphysik, Universit\"at Innsbruck, A-6020
Innsbruck, Austria$^{18}$}
\end{center}\end{sloppypar}
\vspace{2mm}
\begin{sloppypar}
\noindent
E.~Bouhova-Thacker,
C.K.~Bowdery,
A.J.~Finch,
F.~Foster,
G.~Hughes,
R.W.L.~Jones,$^{1}$
M.R.~Pearson,
N.A.~Robertson
\nopagebreak
\begin{center}
\parbox{15.5cm}{\sl\samepage
Department of Physics, University of Lancaster, Lancaster LA1 4YB,
United Kingdom$^{10}$}
\end{center}\end{sloppypar}
\vspace{2mm}
\begin{sloppypar}
\noindent
I.~Giehl,
K.~Jakobs,
K.~Kleinknecht,
G.~Quast,
B.~Renk,
E.~Rohne,
\mbox{H.-G.~Sander},
H.~Wachsmuth,
C.~Zeitnitz
\nopagebreak
\begin{center}
\parbox{15.5cm}{\sl\samepage
Institut f\"ur Physik, Universit\"at Mainz, D-55099 Mainz, Germany$^{16}$}
\end{center}\end{sloppypar}
\vspace{2mm}
\begin{sloppypar}
\noindent
A.~Bonissent,
J.~Carr,
P.~Coyle,
O.~Leroy,
P.~Payre,
D.~Rousseau,
M.~Talby
\nopagebreak
\begin{center}
\parbox{15.5cm}{\sl\samepage
Centre de Physique des Particules, Universit\'e de la M\'editerran\'ee,
IN$^{2}$P$^{3}$-CNRS, F-13288 Marseille, France}
\end{center}\end{sloppypar}
\vspace{2mm}
\begin{sloppypar}
\noindent
M.~Aleppo,
F.~Ragusa
\nopagebreak
\begin{center}
\parbox{15.5cm}{\sl\samepage
Dipartimento di Fisica, Universit\`a di Milano e INFN Sezione di Milano,
I-20133 Milano, Italy}
\end{center}\end{sloppypar}
\vspace{2mm}
\begin{sloppypar}
\noindent
A.~David,
H.~Dietl,
G.~Ganis,$^{26}$
K.~H\"uttmann,
G.~L\"utjens,
C.~Mannert,
W.~M\"anner,
\mbox{H.-G.~Moser},
R.~Settles,
H.~Stenzel,
W.~Wiedenmann,
G.~Wolf
\nopagebreak
\begin{center}
\parbox{15.5cm}{\sl\samepage
Max-Planck-Institut f\"ur Physik, Werner-Heisenberg-Institut,
D-80805 M\"unchen, Germany\footnotemark[16]}
\end{center}\end{sloppypar}
\vspace{2mm}
\begin{sloppypar}
\noindent
J.~Boucrot,$^{1}$
O.~Callot,
M.~Davier,
L.~Duflot,
\mbox{J.-F.~Grivaz},
Ph.~Heusse,
A.~Jacholkowska,$^{22}$
J.~Lefran\c{c}ois,
\mbox{J.-J.~Veillet},
I.~Videau,
C.~Yuan
\nopagebreak
\begin{center}
\parbox{15.5cm}{\sl\samepage
Laboratoire de l'Acc\'el\'erateur Lin\'eaire, Universit\'e de Paris-Sud,
IN$^{2}$P$^{3}$-CNRS, F-91898 Orsay Cedex, France}
\end{center}\end{sloppypar}
\vspace{2mm}
\begin{sloppypar}
\noindent
G.~Bagliesi,
T.~Boccali,
G.~Calderini,
L.~Fo\`{a},
A.~Giammanco,
A.~Giassi,
F.~Ligabue,
A.~Messineo,
F.~Palla,
G.~Sanguinetti,
A.~Sciab\`a,
G.~Sguazzoni,
R.~Tenchini,$^{1}$
A.~Venturi,
P.G.~Verdini
\samepage
\begin{center}
\parbox{15.5cm}{\sl\samepage
Dipartimento di Fisica dell'Universit\`a, INFN Sezione di Pisa,
e Scuola Normale Superiore, I-56010 Pisa, Italy}
\end{center}\end{sloppypar}
\vspace{2mm}
\begin{sloppypar}
\noindent
G.A.~Blair,
G.~Cowan,
M.G.~Green,
T.~Medcalf,
A.~Misiejuk,
J.A.~Strong,
\mbox{P.~Teixeira-Dias},
\mbox{J.H.~von~Wimmersperg-Toeller}
\nopagebreak
\begin{center}
\parbox{15.5cm}{\sl\samepage
Department of Physics, Royal Holloway \& Bedford New College,
University of London, Egham, Surrey TW20 OEX, United Kingdom$^{10}$}
\end{center}\end{sloppypar}
\vspace{2mm}
\begin{sloppypar}
\noindent
R.W.~Clifft,
T.R.~Edgecock,
P.R.~Norton,
I.R.~Tomalin
\nopagebreak
\begin{center}
\parbox{15.5cm}{\sl\samepage
Particle Physics Dept., Rutherford Appleton Laboratory,
Chilton, Didcot, Oxon OX11 OQX, United Kingdom$^{10}$}
\end{center}\end{sloppypar}
\vspace{2mm}
\begin{sloppypar}
\noindent
\mbox{B.~Bloch-Devaux},$^{1}$
P.~Colas,
S.~Emery,
W.~Kozanecki,
E.~Lan\c{c}on,
\mbox{M.-C.~Lemaire},
E.~Locci,
P.~Perez,
J.~Rander,
\mbox{J.-F.~Renardy},
A.~Roussarie,
\mbox{J.-P.~Schuller},
J.~Schwindling,
A.~Trabelsi,$^{21}$
B.~Vallage
\nopagebreak
\begin{center}
\parbox{15.5cm}{\sl\samepage
CEA, DAPNIA/Service de Physique des Particules,
CE-Saclay, F-91191 Gif-sur-Yvette Cedex, France$^{17}$}
\end{center}\end{sloppypar}
\vspace{2mm}
\begin{sloppypar}
\noindent
N.~Konstantinidis,
A.M.~Litke,
G.~Taylor
\nopagebreak
\begin{center}
\parbox{15.5cm}{\sl\samepage
Institute for Particle Physics, University of California at
Santa Cruz, Santa Cruz, CA 95064, USA$^{13}$}
\end{center}\end{sloppypar}
\vspace{2mm}
\begin{sloppypar}
\noindent
C.N.~Booth,
S.~Cartwright,
F.~Combley,
M.~Lehto,
L.F.~Thompson
\nopagebreak
\begin{center}
\parbox{15.5cm}{\sl\samepage
Department of Physics, University of Sheffield, Sheffield S3 7RH,
United Kingdom$^{10}$}
\end{center}\end{sloppypar}
\vspace{2mm}
\begin{sloppypar}
\noindent
K.~Affholderbach,
A.~B\"ohrer,
S.~Brandt,
C.~Grupen,
A.~Ngac,
G.~Prange,
U.~Sieler
\nopagebreak
\begin{center}
\parbox{15.5cm}{\sl\samepage
Fachbereich Physik, Universit\"at Siegen, D-57068 Siegen,
 Germany$^{16}$}
\end{center}\end{sloppypar}
\vspace{2mm}
\begin{sloppypar}
\noindent
G.~Giannini
\nopagebreak
\begin{center}
\parbox{15.5cm}{\sl\samepage
Dipartimento di Fisica, Universit\`a di Trieste e INFN Sezione di Trieste,
I-34127 Trieste, Italy}
\end{center}\end{sloppypar}
\vspace{2mm}
\begin{sloppypar}
\noindent
J.~Rothberg
\nopagebreak
\begin{center}
\parbox{15.5cm}{\sl\samepage
Experimental Elementary Particle Physics, University of Washington, Seattle, 
WA 98195 U.S.A.}
\end{center}\end{sloppypar}
\vspace{2mm}
\begin{sloppypar}
\noindent
S.R.~Armstrong,
K.~Cranmer,
P.~Elmer,
D.P.S.~Ferguson,
Y.~Gao,$^{20}$
S.~Gonz\'{a}lez,
O.J.~Hayes,
H.~Hu,
S.~Jin,
J.~Kile,
P.A.~McNamara III,
J.~Nielsen,
W.~Orejudos,
Y.B.~Pan,
Y.~Saadi,
I.J.~Scott,
J.~Walsh,
Sau~Lan~Wu,
X.~Wu,
G.~Zobernig
\nopagebreak
\begin{center}
\parbox{15.5cm}{\sl\samepage
Department of Physics, University of Wisconsin, Madison, WI 53706,
USA$^{11}$}
\end{center}\end{sloppypar}
}
\footnotetext[1]{Also at CERN, 1211 Geneva 23, Switzerland.}
\footnotetext[2]{Now at Universit\'e de Lausanne, 1015 Lausanne, Switzerland.}
\footnotetext[3]{Also at Dipartimento di Fisica di Catania and INFN Sezione di
 Catania, 95129 Catania, Italy.}
\footnotetext[4]{Deceased.}
\footnotetext[5]{Also Istituto di Cosmo-Geofisica del C.N.R., Torino,
Italy.}
\footnotetext[6]{Supported by the Commission of the European Communities,
contract ERBFMBICT982894.}
\footnotetext[7]{Supported by CICYT, Spain.}
\footnotetext[8]{Supported by the National Science Foundation of China.}
\footnotetext[9]{Supported by the Danish Natural Science Research Council.}
\footnotetext[10]{Supported by the UK Particle Physics and Astronomy Research
Council.}
\footnotetext[11]{Supported by the US Department of Energy, grant
DE-FG0295-ER40896.}
\footnotetext[12]{Now at Departement de Physique Corpusculaire, Universit\'e de
Gen\`eve, 1211 Gen\`eve 4, Switzerland.}
\footnotetext[13]{Supported by the US Department of Energy,
grant DE-FG03-92ER40689.}
\footnotetext[14]{Also at Rutherford Appleton Laboratory, Chilton, Didcot, UK.}
\footnotetext[15]{Permanent address: Universitat de Barcelona, 08208 Barcelona,
Spain.}
\footnotetext[16]{Supported by the Bundesministerium f\"ur Bildung,
Wissenschaft, Forschung und Technologie, Germany.}
\footnotetext[17]{Supported by the Direction des Sciences de la
Mati\`ere, C.E.A.}
\footnotetext[18]{Supported by the Austrian Ministry for Science and Transport.}
\footnotetext[19]{Now at SAP AG, 69185 Walldorf, Germany.}
\footnotetext[20]{Also at Department of Physics, Tsinghua University, Beijing, The People's Republic of China.}
\footnotetext[21]{Now at D\'epartement de Physique, Facult\'e des Sciences de Tunis, 1060 Le Belv\'ed\`ere, Tunisia.}
\footnotetext[22]{Now at Groupe d' Astroparticules de Montpellier, Universit\'e de Montpellier II, 34095 Montpellier, France.}
\footnotetext[23]{Also at Dipartimento di Fisica e Tecnologie Relative, Universit\`a di Palermo, Palermo, Italy.}
\footnotetext[24]{Now at CERN, 1211 Geneva 23, Switzerland.}
\footnotetext[25]{Now at ISN, Institut des Sciences Nucl\'eaires, 53 Av. des Martyrs, 38026 Grenoble, France.} 
\footnotetext[26]{Now at INFN Sezione di Roma II, Dipartimento di Fisica, Universit\'a di Roma Tor Vergata, 00133 Roma, Italy.} 
\footnotetext[27]{Now at LBNL, Berkeley, CA 94720, U.S.A.}
%
\setlength{\parskip}{\saveparskip}
\setlength{\textheight}{\savetextheight}
\setlength{\topmargin}{\savetopmargin}
\setlength{\textwidth}{\savetextwidth}
\setlength{\oddsidemargin}{\saveoddsidemargin}
\setlength{\topsep}{\savetopsep}
\normalsize
\newpage
\pagestyle{plain}
\setcounter{page}{1}